\documentclass[aps,prd,twocolumn,preprintnumbers,superscriptaddress,10pt,nofootinbib]{revtex4-1}
\usepackage{epsf,color,amsmath,amssymb,amsfonts}
\usepackage{enumerate}
\usepackage{hyperref}
\usepackage{graphicx}
\usepackage{psfrag}
\usepackage{dcolumn,bm}

\newcommand{\bea}{\begin{eqnarray}}
\newcommand{\beal}[1]{\begin{eqnarray}\label{#1}}
\newcommand{\eea}{\end{eqnarray}}
\newcommand{\be}{\begin{equation}}
\newcommand{\bel}[1]{\begin{equation}\label{#1}}
\newcommand{\ee}{\end{equation}}

\newcommand{\bit}{\begin{itemize}}
\newcommand{\eit}{\end{itemize}}
\newcommand{\ben}{\begin{enumerate}}
\newcommand{\een}{\end{enumerate}}

\newcommand{\f}[2]{\frac{#1}{#2}}

\begin{document}
\onecolumngrid

\title{Towards a holographic realization of the quarkyonic phase}

\author{Jan de Boer}
\email{j.deboer@uva.nl}
\affiliation{Instituut voor Theoretische Fysica, Universiteit van Amsterdam \\
Science Park 904, 1090 GL Amsterdam, The Netherlands}

\author{Borun D. Chowdhury}
\email{b.d.chowdhury@uva.nl}
\affiliation{Instituut voor Theoretische Fysica, Universiteit van Amsterdam \\
Science Park 904, 1090 GL Amsterdam, The Netherlands}

\author{Michal P. Heller}
\email{m.p.heller@uva.nl}
\altaffiliation[On leave from: ]{\emph{National Centre for Nuclear Research,  Ho{\.z}a 69, 00-681 Warsaw, Poland.}}
\affiliation{Instituut voor Theoretische Fysica, Universiteit van Amsterdam \\
Science Park 904, 1090 GL Amsterdam, The Netherlands}

\author{Jakub Jankowski}
\email{jakubj@ift.uni.wroc.pl}
\affiliation{Institute for Theoretical Physics, University of Wroclaw, 50-204 Wroclaw, Poland}

\begin{abstract}
Large-$N_c$ QCD matter at intermediate baryon density and low temperatures has been conjectured to be in the so-called quarkyonic phase, i.e., to have a quark Fermi surface and on top of it a confined spectrum of excitations. It has been suggested that the presence of the quark Fermi surface leads to a homogeneous phase with restored chiral symmetry, which is unstable towards creating condensates that break both the chiral and translational symmetry. Motivated by these exotic features, we investigate properties of cold baryonic matter in the single-flavor Sakai-Sugimoto model, searching for a holographic realization of the quarkyonic phase. We use a simplified mean-field description and focus on the regime of parametrically large baryon densities, of the order of the square of the 't Hooft coupling, as they turn out to lead to new physical effects similar to the ones occurring in the quarkyonic phase. One effect--the appearance of a particular marginally stable mode breaking translational invariance and linked with the presence of the Chern-Simons term in the flavor-brane Lagrangian--is known to occur in the deconfined phase of the Sakai-Sugimoto model, but turns out to be absent here. The other, completely new phenomenon that we, preliminarily, study using strong simplifying assumptions are density-enhanced interactions of the flavor-brane gauge field with holographically represented baryons. These seem to significantly affect the spectrum of vector and axial mesons and might lead to approximate chiral symmetry restoration in the lowest part of the spectrum, where the mesons start to qualitatively behave like collective excitations of the dense baryonic medium. We discuss the relevance of these effects for holographic searches of the quarkyonic phase and conclude with a discussion of various subtleties involved in constructing a mean-field holographic description of a dense baryonic medium.
\end{abstract}

\maketitle

\section{Introduction}
One of the outstanding open problems in the theory of strong interactions is the phase structure of QCD at low temperatures and at intermediate baryon densities. This region of the phase diagram is relevant for the physics of neutron stars \cite{Haensel:2007yy} and is a subject of significant research efforts in the heavy-ion collision experiments at low and moderate energies \cite{Friman:2011zz}.

In the vacuum and for low temperatures and densities, QCD is realized in the hadronic, confined phase, where the chiral symmetry is broken. Based on lattice QCD simulations \cite{Aoki:2006we,Bazavov:2011nk} and 't Hooft anomaly matching \cite{Coleman:1980mx} we have strong theoretical reasons to suspect that as we increase the temperature at vanishing baryon density, there is a single transition of a crossover nature to a deconfined, chirally symmetric phase--the quark-gluon plasma. 

However, little is known  about the opposite region of zero temperature and intermediate densities, as in this regime of parameters it is hard to apply first-principle methods. More concretely, at moderate temperatures and baryon densities the perturbative expansion faces convergence issues, lattice simulations suffer from the fermionic sign problem, and conclusions from anomaly matching can no longer be drawn because  Lorentz symmetry is broken.

Using qualitative large-$N_c$ arguments, McLerran and Pisarski have recently conjectured that at intermediate baryon density and at low enough temperatures a new, so-called quarkyonic phase appears \cite{McLerran:2007qj}. This phase has a Fermi sea of quarks, but the excitation spectrum consists of color singlets  just as it does in the vacuum. Moreover, quarks from the Fermi sea occupy states that are required to form the chiral condensate, and as a consequence, above a certain critical density the condensate is believed to vanish, leading to chiral symmetry restoration. This is reflected in the spectrum of mesons obtained using model studies in Refs. \cite{Glozman:2007tv,Glozman:2008kn}, which consists of exact chiral multiplets.

In the presence of interactions, the quark Fermi surface is unstable against the formation of quark-quark hole condensates  \cite{Deryagin:1992rw}. This breaks the chiral symmetry again, but now the chiral condensate oscillates in space around its zero value. A preliminary analysis of this phenomenon in the context of quarkyonic matter was done in Refs. \cite{Kojo:2009ha,Kojo:2010fe}, showing that it can have a significant impact on the low-energy behavior of the system.  

A natural consequence of the existence of a new phase is the appearance of a triple point in the QCD phase diagram. This possibility has been explored in order to predict--both qualitatively and semi-quantitatively--the features of hadron production in heavy-ion collisions \cite{Andronic:2009gj}.

During the last ten years a rapid development of  gauge-gravity duality \cite{Maldacena:1997re,Witten:1998qj,Gubser:1998bc} has taken place. It provides a novel first-principle method of solving certain strongly coupled gauge theories in the large-$N_c$ limit. In particular, holographic models of QCD share many features of QCD in the vacuum; for example, there is some similarity in the spectrum of the lowest-lying mesons including the chiral symmetry breaking \cite{Sakai:2004cn}. Another well-known feature where there is approximate agreement are the transport properties (like the famous ratio of the 
shear viscosity to the entropy density \cite{Policastro:2001yc}) in the strongly coupled quark-gluon plasma phase, as observed at the Relativistic Heavy Ion Collider and LHC.
These and other results make the gauge-gravity duality interesting not only from the point of view of QCD phenomenology, but also from the more fundamental perspective of improving our theoretical understanding of nonperturbative phenomena in gauge theories (see Ref. \cite{Gubser:2009md} for a broad overview of these research trends). 

In the gauge-gravity duality there are no fundamental obstacles in studying holographic models at nonzero baryon density, or more generally in the presence of fermionic densities\footnote{The latter observation lies at the very heart of the AdS/CMT program, which tries to apply holography to condensed matter systems (see Ref. \cite{Hartnoll:2011fn} for a recent review).}. Hence, the results obtained using the gauge-gravity duality might also shed light on otherwise inaccessible quantities in cold QCD at moderate baryon densities, as they previously did in the case with the transport properties of the strongly coupled quark-gluon plasma. 

Our approach is based on noting that McLerran and Pisarski's argument about the existence of the quarkyonic phase follows from large-$N_c$ counting and so might also be applicable to holographic models. We hope that first-principle investigations in string theory constructions will bring new results to the existing phenomenology of the quarkyonic phase and also lead to further progress in understanding holographic QCD. 

There are three issues that seem to be crucial for claiming a holographic realization of the quarkyonic phase. We address all of these at a preliminary level in our paper: forming a quark Fermi surface in the holographic model of interest when it is put at finite baryon density, having colorless excitations of the quark Fermi surface with the chiral symmetry (approximately) restored in the spectrum, and identifying a possible mechanism leading to instabilities breaking translational invariance at large distances.

As a holographic QCD model we choose the one introduced by Sakai and Sugimoto in Refs. \cite{Sakai:2004cn,Sakai:2005yt} and study its properties at vanishing temperature but nonzero baryon density. We focus on the case of having a single quark flavor ($N_{f} = 1$). Our chief motivation for focusing on this particular realization of holographic QCD is its simplicity and the fact that it nevertheless incorporates chiral symmetry breaking. In the large-$N_{c}$ limit, the $N_{f} = 1$ analogue of the chiral symmetry breaking is the axial symmetry breaking, and we will refer to this phenomenon simply as the chiral symmetry breaking.

The model of interest \cite{Sakai:2004cn,Sakai:2005yt} is based on a system of $N_c$ D4-branes compactified on a circle with antiperiodic boundary conditions for fermions. Upon backreaction these generate a spacetime dual to the adjoint sector of the (3+1)-dimensional gauge theory \cite{Witten:1998zw}. $N_f$ D8-branes and $N_{f}$ $\overline{\mathrm{D8}}$-branes localized at two different points of the circle  represent $N_f$ flavors of chiral quarks in the quenched approximation \cite{Karch:2002sh}. 

At vanishing temperature--which is what we are interested in--the spacelike circle shrinks to zero size in the IR part of the geometry. The D8- and $\overline{\mathrm{D8}}$-branes, which in the UV are localized at different points on the circle, must then connect in the IR and form a U-shaped configuration. This purely geometric mechanism is the dual description of the chiral symmetry breaking. Indeed, normalizable solutions of the gauge field on D8-branes in vacuum--corresponding to some of the spin-0 and spin-1 mesons--have $N_f^2$ massless modes interpreted as $N_f^2 - 1$ pions and the $\eta'$ meson, which in the large-$N_c$ limit becomes the Goldstone mode of the spontaneously broken axial symmetry \cite{Witten:1979vv}. Another indication of chiral symmetry breaking is that the  holographically obtained masses of the axial and vector mesons differ.

In the Sakai-Sugimoto model, a nonzero baryon density at vanishing temperature is realized by an antisymmetric configuration of the electric field in the holographic direction on the U-shaped D8-brane. Such a configuration requires sources (baryons), which in the Abelian case considered here are D4-branes living on the D8-brane and wrapping the $t \times S^4$ part of the geometry. After a Kaluza-Klein reduction on $S^4$, the D4-branes correspond to massive, electrically charged point particles living in the world volume of the flavor D8-brane.

The model studies of the quarkyonic phase suggest that the chiral symmetry gets restored at sufficiently large baryon densities \cite{Forkel:1989wc,Glozman:2007tv,Glozman:2008kn}. On the gravity side there is no obvious way of achieving chiral symmetry restoration in the confining phase, as opposed to the deconfined phase, where there is a natural mechanism to decouple the D8- and $\overline{\mathrm{D8}}$-branes by extending them all the way to the horizon. Therefore, if there is a dual description of the quarkyonic phase it will necessarily involve some new ingredients.

Although the holographic baryons are point-like (single-flavor case) or almost point-like (two flavors \cite{Hata:2007mb}) objects, on the field theory side they correspond to particles of a finite size \cite{Hashimoto:2008zw} when both the number of colors $N_{c}$ and the 't Hooft coupling constant $\lambda$ are very large. This seems to suggest that in order to have overlapping baryons on the field theory side, and hence at least naively form a quark Fermi surface, one only needs an ${\cal O}{(1)}$ (in both $N_{c}$ and $\lambda$) three-dimensional density of point-like holographic baryons residing at the bottom of the U-shaped D8-brane. Attempts to construct such holographic baryon densities were made in Refs. \cite{Bergman:2007wp,Kim:2007zm}. We, however, will focus on parametrically higher densities, of order of ${\cal O}{(\lambda^{2})}$, as \emph{only} in this case is there a possibility of having new physical effects bearing similarities to the ones expected to occur in the quarkyonic phase. The first such effect--possible instabilities towards breaking translational invariance triggered by the presence of the Chern-Simons term in the flavor-brane Lagrangian--is already present in the literature on holographic QCD \cite{Kim:2010pu,Ooguri:2010xs,Bayona:2011ab} but has never been comprehensively analyzed at vanishing temperature and nonzero baryon density. The second--to the best of our knowledge--completely new effect that we investigate here are density-enhanced interactions of mesons with baryons.  As we show later in the article, these  may significantly affect the meson spectrum.

It turns out that such parametrically large baryon densities have a peculiar effect on the distribution of the holographic baryons (D4-branes wrapping $S^{4}$) on the flavor brane. In the discussion below we will follow Ref. \cite{Rozali:2007rx}, where it was noticed that due to the electromagnetic repulsion, at densities of order ${\cal O}{(\lambda^{2})}$ the D4-branes will fill not only the field theory directions but also the holographic direction. Such a configuration was explicitly constructed in Ref. \cite{Rozali:2007rx} in a mean-field approach for the antipodal embedding of the D8-brane, and it is the starting point for our considerations here. A further justification that the holographic baryon density will fill four spatial directions, including the holographic direction, at a large enough baryon density on the boundary comes from the recent study in Ref.\cite{Kaplunovsky:2012gb}. There the authors--based on large-$N_{c}$ scaling--argued that the structure of the ground state of this holographic model of QCD is necessarily a crystal on the flavor brane. Such a holographic crystal structure for sufficiently large densities will span not only the field theory but also the holographic directions, as energetic arguments show.

As mentioned earlier, mesons appear as normalizable perturbations of the flavor-brane gauge field. It is natural to expect that such modes persist at nonzero baryon density, albeit with possibly different properties than in the vacuum. We already see at this level that in the holographic model we can have a quark Fermi surface--assuming having overlapping baryons and strong interactions between their holographic counterparts is sufficient for this--and colorless excitations at the same time. These are the defining features of the quarkyonic phase.

A generic feature of holographic QCD models is the presence of a Chern-Simons term for the gauge field which represents the chiral flavor degrees of freedom in the bulk, and which is responsible for the flavor anomaly on the field theory side \cite{Witten:1998qj,Domokos:2007kt}. Recently, it has been shown that the presence of the Chern-Simons term together with a nonzero value of the bulk electric field may generate unstable modes carrying nonzero momentum \cite{Domokos:2007kt,Nakamura:2009tf}. The condensation of such modes leads to the appearance of an inhomogeneous phase \cite{Ooguri:2010kt}. Indeed, such an effect is present in the Sakai-Sugimoto model in the deconfined phase at nonzero baryon density \cite{Ooguri:2010xs} and also at vanishing temperature and sufficiently large axial chemical potential \cite{Bayona:2011ab}. 

Although the instability pattern resembles the one found in the case of the quarkyonic phase, the operators that are condensing are different. Superficially, this seems to be enough to discourage linking Chern-Simons term-induced instabilities with the ones appearing in the quarkyonic phase \cite{Kojo:2009ha,Kojo:2010fe}. However, it is  conceivable that the instability due to the Chern-Simons term also triggers a modulation in the operators, which are conjectured to condense in the quarkyonic phase. We will not pursue this exciting idea here besides some brief comments in the concluding remarks, but we will nevertheless search for modulated phases due to the Chern-Simons coupling, as the breaking of translational invariance will affect the properties of the background solution and hence the spectrum of mesons.  We will follow Ref. \cite{Ooguri:2010xs} to search  for an indication of an instability and  for marginally stable modes, i.e., time-independent normalizable solutions of the equations of motion for the bulk gauge field carrying nonzero momentum. It is interesting to note that the Chern-Simons term, which is often crucial in order to obtain instabilities, becomes relevant only for large densities--like the ones that we consider here--and otherwise its influence on the dynamics of the theory is subleading at \mbox{large $\lambda$.}

In this paper we take a first step in trying to understand the so-far neglected interactions of mesons with the dense baryonic medium. This seems to be the simplest possible mechanism leading to a partial decoupling between the D8- and $\overline{\mathrm{D8}}$-brane in the regime of interest. In hindsight, the mean-field treatment of the $N_f = 1$ configuration originally adopted in Ref. \cite{Rozali:2007rx} turns out to be a brane analogue of an electron star system introduced in the AdS/CMT context in Ref. \cite{Hartnoll:2010gu}. The relevant action  is that of a charged dust composed of D4-branes wrapping the four-sphere and residing on the D8-brane, coupled to the D8-brane Dirac-Born-Infeld (DBI) action for the electromagnetic field. We argue that such an action captures part of the response of the baryonic medium to meson perturbations. The element that is not taken into account by the dust approximation is a restoring force acting on the holographic baryons once they are perturbed from their equilibrium position in the underlying crystal lattice.

We leave the study of this (possibly very important yet hard to implement in the coarse-grained description) effect for future investigations. We subsequently focus on finding-zero momentum transverse fluctuations of the gauge field (corresponding to spin-one mesons) in the presence of a nonzero baryon density and show that the electromagnetic interactions with the D4-branes on the D8-brane significantly affect the frequencies (meson masses). In particular, the lowest mesons at large densities turn out to be  bound states in an effective potential generated by the presence of the baryonic charge on the flavor brane. This leads to the possibility that the spectrum organizes itself in approximate axial (the $N_{f} = 1$ analogue of chiral) doublets. In particular, the analysis of this effective potential--which appears in the Schr\"{o}dinger form of the equation for in-medium meson perturbations--suggests that such a possibility may indeed be realized for a particular form of the embedding, albeit only for the lowest axial and vector meson. We comment on the relevance of the restoring force that we neglected in the Conclusions section.

The plan of the paper is as follows. In Sec. \ref{sec.holomodel} we review various features of the Sakai-Sugimoto model relevant for our studies and explain how to incorporate baryons in the holographic description. In Sec. \ref{sec.motivationforlargedensity} we show how the coupling-constant ($\lambda$) dependence of different terms in the bulk Lagrangian leads to possibly interesting physical effects at baryon densities that scale as ${\cal O}{(\lambda^{2})}$. In Sec. \ref{sec.meanfieldattempt} we attempt to construct zero-temperature states in the $N_f$ = 1 Sakai-Sugimoto model with large baryon density for antipodal and nonantipodal embeddings, generalizing the results of Ref. \cite{Rozali:2007rx}. In Secs. \ref{sec.stability} and \ref{sec.mesons} we discuss interesting physical effects associated with large densities. In Sec. \ref{sec.stability} we search for normalizable time-independent perturbations of a gauge field carrying nonzero momentum, whose presence signals an instability towards creating striped structures. Section \ref{sec.mesons} is devoted to recovering part of the spectrum of mesons at nonzero density and explaining a possible mechanism of chiral symmetry restoration. In the last section we summarize our results and discuss open problems.

\section{The holographic model \label{sec.holomodel}}

\subsection{Vacuum solution \label{subsec.vacuumsol}}

The starting point of our studies is the geometry generated by $N_c$ D4-branes compactified on a circle with antiperiodic boundary conditions for the (adjoint) fermions. Such a background, representing the ground state of the adjoint sector in the dual field theory, is a solution of type IIA supergravity with a nontrivial metric $g_{AB}$, the dilaton $\phi$ and the four-form $F_4$. In the following we will denote the noncompact D4 world-volume (``field theory'') directions by $x^{\mu}$, the coordinate on the compactification circle of circumference $\delta \tau$ by $\tau$,
\be
\tau \equiv \tau + \delta \tau,
\ee
the radial (``holographic'') direction by $u$, the line element on the four-sphere by $d\Omega_{4}^{2}$, the unit four-sphere volume by $V_{4}$
\be
V_{4} = \frac{8}{3} \pi^2 
\ee
and its volume form by $\epsilon_4$. The full solution now reads
\bea
&&ds^2 = g_{A B} dx^{A} dx^{B} = (u/R)^\f{3}{2} ( \eta_{\mu \nu} dx^{\mu}dx^{\nu}+ f(u) d\tau^2)+\nonumber\\ &&+ (u/R)^{-\f{3}{2}} (\f{du^2}{f(u)} + u^2 d\Omega_4^2), \nonumber \\
&&e^\Phi= g_s (u/R)^\f{3}{4} \quad \mathrm{and} \quad F_4 =  \f{2\pi N_c}{V_4} \epsilon_4, \label{eq.gphiF4}
\eea
with the curvature scale set by $R$, the string length and  the string coupling denoted respectively by $l_{s}$ and $g_{s}$, and $f(u)$ being
\be
f(u) = 1 - \left(\frac{u_{KK}}{u}\right)^3.
\ee
Regularity at the tip of the cigar formed by the $u$ and $\tau$ directions relates $u_{KK}$ and $\delta \tau$,
\be
\delta \tau = \frac{4\pi}{3} \frac{R^{3/2}}{u_{KK}^{1/2}}.
\ee
After introducing the Kaluza-Klein mass
\be
M_{KK} = \frac{2\pi}{\delta \tau} = \frac{3}{2} \frac{u_{KK}^{1/2}}{R^{3/2}},
\ee
the mapping between the field theory parameters--the number of colors $N_c$ and the 't Hooft coupling constant $\lambda = g_{YM}^{2} N_{c}$--and $R$, $l_{s}$, and $g_s$  becomes
\be
g_{s}^2 = \frac{1}{8\pi^2} \frac{\lambda^3}{M_{KK}^3 R^3 N_c^2} \quad \mathrm{and} \quad l_s^2 = 2 M_{KK} R^{3} \, \lambda^{-1}.
\ee
Note that everywhere in the text we are using the \emph{four-dimensional} 't Hooft coupling constant $\lambda$ and the \emph{four-dimensional} Yang-Mills coupling constant $g_{YM}$. It is perhaps worth clarifying here the conditions under which we can trust the supergravity approximation. As explained in detail in Ref. \cite{Kruczenski:2003uq}, demanding the smallness of $g_{s}$ and $l_{s}/R$ while keeping $M_{KK} R$ fixed amounts to
\be
1\ll g_{YM}^2 N_c \ll 1/g_{YM}^4,
\ee
which in terms of the 't Hooft coupling constant $\lambda$ and the number of colors $N_{c}$ translates to the following set of inequalities\footnote{In particular, in the remainder of this paper, when we discuss the scaling of various quantities as $\lambda\rightarrow \infty$ we implicitly assume that we keep $M_{KK}R$ fixed and $g_{s}$ fixed and very small, and at the same time we insist on obeying the inequality \eqref{eq.Ncineq}.}:
\be
\label{eq.Ncineq}
1 \ll \lambda^{3/2} \ll N_{c}.
\ee

The key insight brought by Sakai and Sugimoto is the realization that, from the perspective of the dual field theory, adding to this background $N_f$ coincident probe D8-branes and $N_f$ coincident probe $\overline{\mathrm{D8}}$-branes localized  at different points of the circular direction $\tau$  at large $u$, introduces $N_f$ chiral flavors in the quenched approximation, and also leads to the chiral symmetry breaking in the IR. Indeed, the form of the background geometry necessitates that asymptotically separated D8- and $\overline{\mathrm{D8}}$-branes must connect at some smaller value of the radial coordinate and form a U-shaped stack. In the following we will focus on the $N_f = 1$ case and we will consider the asymptotic separation of D8- and $\overline{\mathrm{D8}}$, denoted as $L \leq \delta \tau /2$, as an additional independent parameter specifying the model of holographic QCD. 

To summarize, dual field theory has two dimensionless parameters [$N_c$ and $\lambda$ (as $N_f = 1$)] and two scales ($M_{KK}$ and $L$). Although $M_{KK}$ plays the role of $\Lambda_{QCD}$ and in particular sets the scale for glueball masses, $L$ seems to have no analogue in QCD.

The dynamics of the flavor degrees of freedom is governed by the DBI action for, in our case, a single D8-brane. This object wraps the $x^{\mu}$ and the $u$ directions, as well as the four-sphere. The degrees of freedom appearing in the action are ten embedding functions $X^{A}$ of nine world-volume coordinates $\sigma^{a}$ and the U(1) gauge field $A_{a}$, which has the field strength $F_{a b}$. With the induced metric given by
\be
g_{a b} = g_{AB} \partial_{a} X^{A} \partial_{b} X^{B},
\ee
the DBI action reads
\be
\label{eq.SDBI}
S_{DBI} = - \mu_{8} \int d^{9} \sigma e^{-\phi} \sqrt{- \mathrm{det}\left( g_{a b} + 2 \pi l_{s}^{2} F_{a b} \right)},
\ee
where
\be
\mu_{8} = \frac{1}{(2 \pi)^{8} l_{s}^{9}}.
\ee
The action \eqref{eq.SDBI} needs to be supplemented by the coupling of the U(1) gauge field to the background $C_3$ form  given by 
\be
\label{eq.SCS}
S_{CS} = \frac{1}{6} \mu_{8} \int_{D8} \, C_{3} \wedge \mathrm{Tr} (2\pi l_{s}^{2} F)^{3}.
\ee
In the following, we will choose the gauge such that
\be
x^{a} = X^{a}(\sigma) = \sigma^{a}
\ee
for all $X^{a}$ apart from $\tau$, which is the transverse position of the D8-brane. The latter is given by a nontrivial scalar function $y$ of world-sheet coordinates $x$,
\be
\tau = y(x).
\ee
In this article we will be interested in SO(5) singlet configurations, and hence in the Kaluza-Klein reduction on the four-sphere we will keep the zero modes only. After the Kaluza-Klein reduction and integrating by parts, $S_{CS}$ becomes the action for the five-dimensional Abelian Chern-Simons term
\be
\label{e}
S_{CS} = \frac{N_{c}}{24 \pi^2} \int_{\mathbb{R}^{1,3} \times \mathbb{R}_{+}} A \wedge F \wedge F,
\ee
where $\mathbb{R}_{+}$ and $\mathbb{R}^{1,3}$ denote the holographic and the field theory directions, respectively.

Before we move on, it is convenient to rescale the world-volume coordinates and the gauge field by
\bea
\label{eq.rescalings1}
u = R \, \tilde{u}, \quad x^{\mu} = R \, \tilde{x}^{\mu}, \quad y = R \, \tilde{y},  \nonumber\\
g_{a b} = \tilde{g}_{a b} \quad \mathrm{and} \quad A_{a} = \frac{R}{2 \pi l_{s}^{2}} \tilde{A}_{a}.
\eea
Note that rescaling $A_{a}$ as above and treating $\tilde{A}_{a}$ as an ${\cal O}{(1)}$ quantity leads to parametrically large (in $\lambda$) baryon and axial densities. For the latter baryon-baryon and baryon-meson interactions become important. Such rescalings were (implicitly) also made in many of the previous works on holographic QCD at nonzero baryon density, such as Refs. \cite{Rozali:2007rx,Bergman:2007wp,Ooguri:2010xs}. We will come back to this issue in Sec. \ref{sec.motivationforlargedensity}. 

Having done the rescalings \eqref{eq.rescalings1}, the five-dimensional action reads
\bea
\label{eq.Seff}
S/c =&& - \int d\tilde{u} \, d^{4}\tilde{x} \,\, \tilde{u}^{1/4} \sqrt{- \det{(\tilde{g}_{ab} + \tilde{F}_{a b})}} \nonumber\\
&&+\alpha \int  d\tilde{u} \, d^{4}\tilde{x} \,\, \epsilon_{a b c d e} \tilde{A}^{a} \tilde{F}^{b c} \tilde{F}^{d e},
\eea
where $\epsilon_{a b c d e}$ is the Levi-Civita symbol, the overall dimensionless factor $c$ is given by
\be
\label{eq.cbulk}
c = \mu_{8} V_{4} R^{9} g_{s}^{-1},
\ee
or in terms of the field theory parameters by
\be
\label{eq.cqft}
c = \frac{1}{768 \pi^{5} R^3 M_{KK}^{3}} N_{c} \lambda^3
\ee
and the coefficient of the Chern-Simons term $\alpha$ reads
\be
\alpha = \frac{1}{8}.
\ee

\begin{figure}
\includegraphics[width=8.5cm]{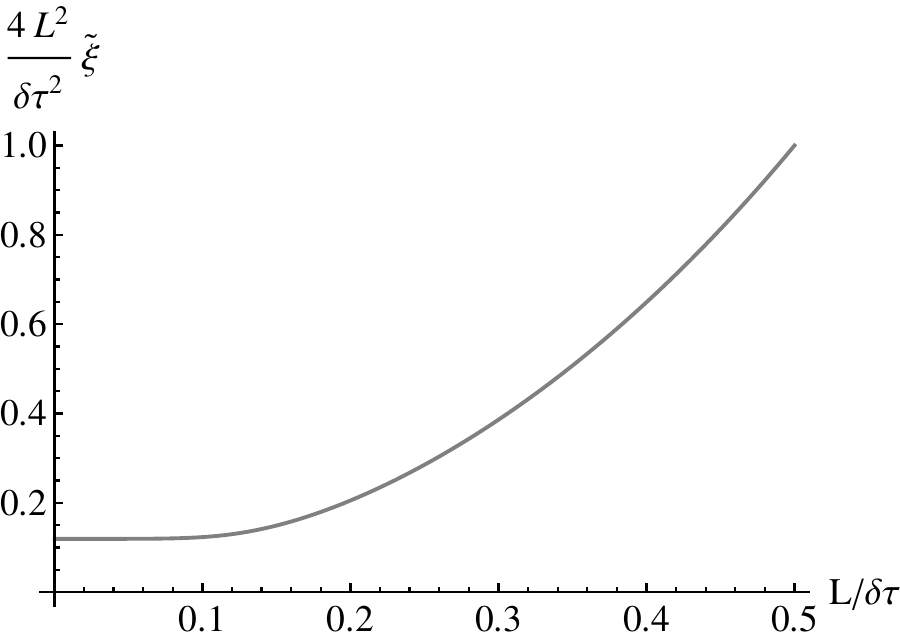}
\caption{The plot of the lowest position of the flavor brane $\xi$, as a function of the asymptotic separation $L$. The ratio $L/\delta \tau = 0.5$ corresponds to  antipodal embedding.}
\label{fig.asymptotic_separation_in_vacuum}
\end{figure}

In the vacuum $\tilde{F}_{ab}$ vanishes and the embedding of D8-brane is given by $\tau$, which is a function of the radial coordinate only. The latter is also true at nonzero baryon density. As the D8-brane is shaped like the letter ``U'' and has two asymptotic regions, we will define the world-volume radial coordinate $\tilde{z}$ as
\be
\tilde{u}/\tilde{u}_{KK} = \tilde{\xi} \left(1+\frac{\tilde{z}^2}{\tilde{\xi}^2}\right)^{\frac{1}{3}}.
\ee
In this way, running $\tilde{z}$ from $-\infty$ to $\infty$ interpolates between two asymptotic regions of the D8-brane, and $\tilde{\xi}$ denotes the lowest radial position of the D8-brane world volume. For antipodal embedding, i.e. for the D8- and $\overline{\mathrm{D8}}$-brane asymptotically localized at opposite points of the circle ($L = \delta \tau /2$), symmetry dictates that $\tilde{\xi} = 1$ and $\tilde{y}'(\tilde{z}) = 0$. In more general situations, $\tilde{\xi}$ will be a nontrivial function of $L$, determined by solving the equations of motion for the D8-brane embedding function $\tilde{y}(\tilde{z})$
\be
\label{eq.Lintermsofy}
\tau/R = \tilde{y}(\tilde{z}) \quad \mathrm{subject \,\, to} \quad L/R = 2 \int_{0}^{\infty} \tilde{y}'(\tilde{z}) d\tilde{z}.
\ee
For $\tilde{y}$ as a function of $\tilde{z}$ only, the induced metric on $\mathbb{R}^{1,3} \times \mathbb{R}_{+}$ reads
\bea
\label{eq.inducedmetric}
&&\tilde{g}_{a b} d\tilde{x}^{a} d\tilde{x}^{b} = \Big\{ \frac{4 \tilde{\xi}^{7/6} \tilde{z}^2}{9 (\tilde{\xi}^2 + \tilde{z}^2)^{5/6} (\tilde{\xi}^{3}-1+\tilde{z}^{2}\tilde{\xi})} \\&&+ \frac{(\tilde{\xi}^{3}-1+\tilde{z}^{2}\tilde{\xi}) \tilde{y}'(\tilde{z})^{2}}{\sqrt{\tilde{\xi}(\tilde{\xi}^2 + \tilde{z}^2)}} \Big\} \, d\tilde{z}^2 + \sqrt{\tilde{\xi}(\tilde{\xi}^2 + \tilde{z}^2)}\,\eta_{\mu \nu} d\tilde{x}^{\mu} d\tilde{x}^{\nu} \nonumber.
\eea
The equations of motion for $\tilde{y}(\tilde{z})$ have the first integral which, for vanishing $\tilde{F}_{a b}$, takes the form
\be
\label{eq.yz}
\frac{\tilde{\xi} ^{9/4} \big(1+\frac{\tilde{z}^2}{\tilde{\xi}^2}\big)^{19/12} \left(\tilde{\xi}^3+\tilde{\xi} \tilde{z}^2-1\right) \tilde{y}'(\tilde{z})}
{\sqrt{-\mathrm{det}(\tilde{g}_{ab})}} = \sqrt{\tilde{\xi}^3 - 1}
\ee
and allows us to algebraically solve for $\tilde{y}'(\tilde{z})$. The integration constant on the rhs is fixed by the small-$\tilde{z}$ expansion of the lhs. Similar reasoning will be used to determine the embedding in the presence of very large baryon density. See Fig. \ref{fig.asymptotic_separation_in_vacuum} for a functional dependence of $\tilde{\xi}$ on the asymptotic separation $L$.

In the rest of this article we will (partially) fix the gauge freedom in $\tilde{A}_{a}$ by demanding that $\tilde{A}_{\tilde{z}} = 0$, leaving the $\mu$ components nontrivial.

Small normalizable perturbations of the embedding function $\delta \tilde{y}$ and the gauge field $\delta \tilde{A}_{\mu}$ have the interpretation of spin-zero (scalar and pseudoscalar) and spin-one (vector and pseudovector, here referred to as axial) mesons \cite{Sakai:2004cn}. In this article we focus our attention only on the latter, which correspond to perturbations $\delta \tilde{A}_{\mu}$ vanishing as $\tilde{z} \rightarrow \pm \infty$. Such an assumption misses a single normalizable (in the sense of having $\delta \tilde{F}_{a b} \rightarrow 0$ as $\tilde{z} \rightarrow \pm \infty$) mode having the asymptotics of a pure four-dimensional gauge, i.e. $\tilde{A}_{\mu} \rightarrow \pm \partial_{\mu} \phi(x^{\nu})$ as $\tilde{z} \rightarrow \pm \infty$. This mode is massless and corresponds to the $\eta'$ meson being, in the large-$N_c$ limit, the Goldstone boson of the spontaneously broken axial symmetry \cite{Sakai:2004cn}.

The masses of the vector and the axial vector mesons can be obtained by analyzing the spectrum of the normalizable perturbations of the form 
\be
\label{eq.mesonperturbation}
\delta \tilde{A}_{\mu} d \tilde{x}^{\mu} = \delta \tilde{a}_{3}(\tilde{z}) e^{- \imath \tilde{\omega} \tilde{t}} d \tilde{x}^{3},
\ee
where the relations between frequencies of normalizable modes $\tilde{\omega}$ and the masses of the mesons $m$ in the dual field theory reads \cite{Sakai:2004cn}
\be
\label{eq.massintermsofomega}
m = \frac{2}{3} M_{KK} \tilde{\omega}.
\ee
Such an ansatz leads to mesons at zero momentum with polarizations in one of the spacelike field theory directions, which without the loss of generality can be chosen to be $\tilde{x}^{3}$. Linearizing the equations of motion for the action \eqref{eq.Seff} leads to the Schr\"odinger-like equation for $\delta \tilde{a}_{3}'(\tilde{z})$ and $\tilde{\omega}$
\bea
\label{eq.vacuummesons}
&&\frac{1}{\sqrt{-\mathrm{det}\,\tilde{g}_{ab}}} \, \frac{d}{d\tilde{z}} \left\{\tilde{\xi}^{19/12}(\tilde{\xi}^2+\tilde{z}^2)^{19/12} \sqrt{-\mathrm{det}\,\tilde{g}_{ab}} \,\, \delta \tilde{a}_{3}'(\tilde{z}) \right\} + \nonumber \\
&&+ \frac{\tilde{\omega}^2}{\tilde{\xi}^{11/12} (\tilde{\xi}^2 + \tilde{z}^2)^{11/12}} \delta \tilde{a}_{3}'(\tilde{z}) = 0,
\eea
where $\mathrm{det}\,\tilde{g}_{ab}$ is the determinant of the induced metric \eqref{eq.inducedmetric} which encodes the dependence on the embedding. Imposing the normalizability condition and identifying the frequency of the eigenmode with the mass of the corresponding meson leads to the meson spectrum in the vector and axial vector channels. As Eq. \eqref{eq.vacuummesons} is invariant under $\tilde{z} \rightarrow - \tilde{z}$, the solutions $\delta \tilde{a}_{3}$ are either symmetric or antisymmetric functions of $\tilde{z}$. As explained in Ref. \cite{Sakai:2004cn}, these correspond respectively to vector (symmetric under  $\tilde{z} \rightarrow - \tilde{z}$) and axial (antisymmetric under  $\tilde{z} \rightarrow - \tilde{z}$) mesons. 

\begin{figure}
\includegraphics[width=8.5cm]{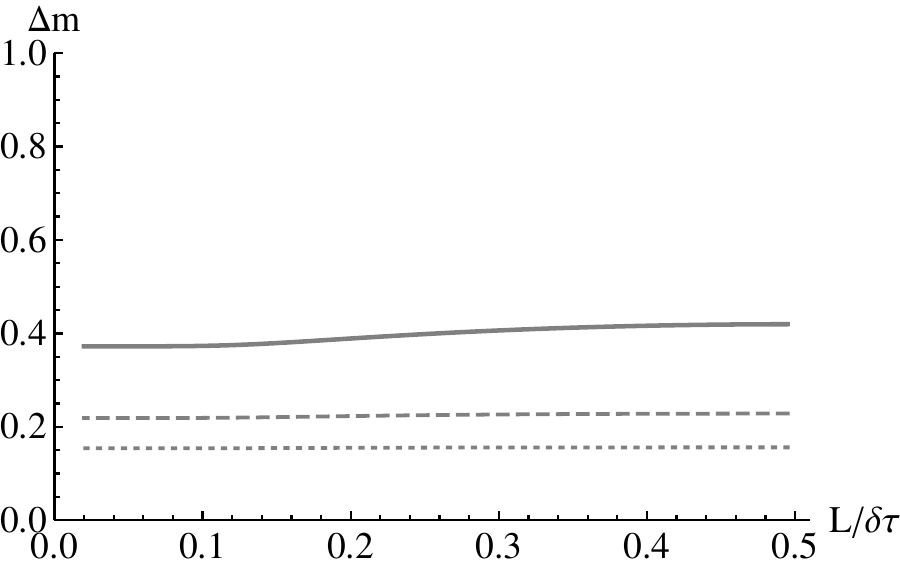}
\caption{The relative mass difference, given by Eq. \eqref{eq.Deltam}, for the three lowest axial and vector mesons. The continuous curve corresponds to the lowest axial and vector mesons, the dashed curve to the second lowest, and the dotted one to the third lowest. The difference is attributed to the axial symmetry breaking, which is the $N_{f} = 1$ analogue of  chiral symmetry breaking.}
\label{fig.Deltam}
\end{figure}

\begin{figure}
\includegraphics[width=8.5cm]{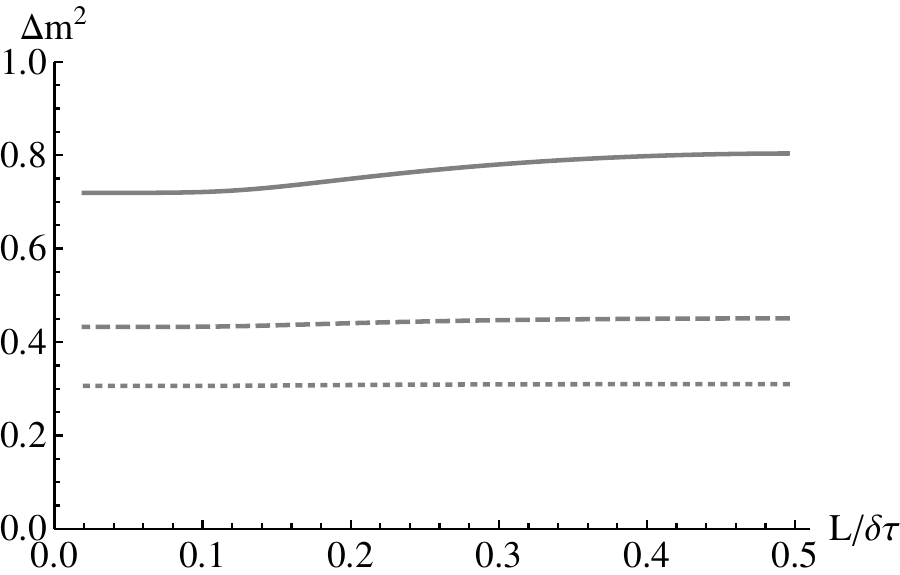}
\caption{The relative mass-squared difference, given by Eq. \eqref{eq.Deltamsq}, for the three lowest axial and vector mesons. The notation is as in \mbox{Fig. \ref{fig.Deltam}}.}
\label{fig.DeltamSq}
\end{figure}

The difference in the masses of the corresponding vector and axial mesons, i.e. the lowest vector and lowest axial mesons, second lowest vector and second lowest axial mesons, etc, is attributed to the axial symmetry breaking. Figures \ref{fig.Deltam} and \ref{fig.DeltamSq} show the relative mass ratios,
\be
\label{eq.Deltam}
\Delta m = 2 \frac{m_{A} - m_{V}}{m_{A} + m_{V}}
\ee
and relative mass-squared ratios,
\be
\label{eq.Deltamsq}
\Delta m^2 = 2 \frac{m_{A}^2 - m_{V}^2}{m_{A}^2 + m_{V}^2},
\ee
for the lowest three vector (with mass $m_{V}$) and corresponding axial mesons (with mass $m_{A}$) as a function of the asymptotic separation between the $\overline{D8}$- and D8-brane $L$. The latter quantities (squares of masses) will become important in Sec. \ref{subsec.SchrodingerPotential},
where we analyze the equation for vacuum meson perturbations \eqref{eq.vacuummesons} by putting it in the form of the Schr\"{o}dinger equation, and also consider its generalization to large densities. The key observation following from Fig. \ref{fig.Deltam} is that the relative mass difference \eqref{eq.Deltam} between the lowest axial and vector mesons is very significant--approximately $40\%$--and does not depend strongly on the asymptotic separation $L$. For higher-order pairs of vector and axial mesons the relative mass difference becomes smaller and for very large quantum numbers becomes insignificant. This can be understood as the chiral symmetry restoration in the UV part of the spectrum \cite{Glozman:2007ek}.

In Sec. \ref{sec.mesons} we will revisit perturbations of the form \eqref{eq.mesonperturbation} in the presence of a parametrically large baryon density and investigate how \emph{some} of the effects generated by the dense baryonic medium affect the masses of vector and axial vector mesons. It will turn out that the  effects considered generically lower the relative mass difference between the mesons in question [Eq. \eqref{eq.Deltam}], in certain situations to as much as $10\%$. We will perform a thorough analysis of this phenomenon in Sec. \ref{subsec.SchrodingerPotential}.

\subsection{Introducing baryons}

Our goal is to study the theory at nonzero baryon density. In the grand canonical ensemble this amounts to modifying its Lagrangian density by
\be
\label{eq.grandcanonicalensamble}
{\cal L} \to {\cal L} - \mu Q
\ee
where $\mu$ is the quark chemical potential and $Q$ is the density of the quark charge expressed in terms of left and right quark operators as
\be
\label{eq.Bdef}
Q =\psi_{L}^\dagger \psi_{L} + \psi_{R}^{\dagger} \psi_{R}.
\ee
In this formulation the baryon density $B$ can be obtained from the quark density $Q$ by
\be
Q = N_{c} B.
\ee
On the gravity side in the gauge $\tilde{A}_{\tilde{z}} = 0$, the fields dual to $\overline{\psi}_{L} \gamma_{\mu} \psi_{L}$ and $\overline{\psi}_{R} \gamma_{\mu} \psi_{R}$ are $\tilde{A}_{\mu}$ components of the $\overline{\mathrm{D8}}$- and D8-brane U(1) gauge field. In other words $\overline{\psi}_{L} \gamma_{\mu} \psi_{L}$ and $\overline{\psi}_{R} \gamma_{\mu} \psi_{R}$ can be obtained respectively from $\tilde{z} = -\infty$ and $\tilde{z} = \infty$ asymptotics of $\tilde{A}_{\mu}(\tilde{z})$. More concretely, the asymptotic solution for the timelike component of the gauge field takes the form
\bea
\tilde{A}_{0}\big|_{\tilde{z}\rightarrow -\infty} &&= \tilde{C}_{L} + \frac{2 \tilde{E}_{L}}{3\sqrt{\tilde{\xi}} \tilde{z}} + \ldots \nonumber\\
\mathrm{and}&&\nonumber\\
\tilde{A}_{0}\big|_{\tilde{z}\rightarrow \infty} \quad &&= \tilde{C}_{R} - \frac{2 \tilde{E}_{R}}{3\sqrt{\tilde{\xi}} \tilde{z}} + \ldots
\eea
where $\tilde{C}_{L/R}$ is related to the left or right quark chemical potential and $\tilde{E}_{L/R}$--the dimensionless electric flux on D8- and $\overline{\mathrm{D8}}$-brane, respectively--is proportional to the density of left or right quarks $q_{L/R}$ in the dual field theory \cite{Rozali:2007rx},
\be
q_{L/R} = \frac{2\pi c \, l_{s}^{2}}{R^{5}} \tilde{E}_{L/R}.
\ee
The latter expression can also be written in terms of the field theory parameters as
\be
\label{eq.qLRintermsoffieldtheoryparams}
q_{L/R} = \frac{N_{c} \lambda^2}{192 \pi^4 M_{KK}^2 R^{5}} \tilde{E}_{L/R}.
\ee
As the baryon density $B$ is a direct sum of $q_{L} = \langle \psi_{L}^\dagger \psi_{L} \rangle$ and $q_{R} = \langle \psi_{R}^{\dagger} \psi_{R}\rangle$, weighted by the factor $1/N_{c}$, achieving a nonzero baryon density at a vanishing axial density ($q_{A} = q_{L} - q_{R}$) requires turning on a symmetric (under $\tilde{z} \leftrightarrow -\tilde{z}$) configuration of the $\tilde{A}_{0}(\tilde{z})$ gauge field. Such configurations give rise to an antisymmetric electric field, which has a nonzero asymptotic flux equal to the sum of the fluxes on the D8- and $\overline{\mathrm{D8}}$-brane parts of the world volume. It is clear from Eq. \eqref{eq.qLRintermsoffieldtheoryparams} that keeping $\tilde{E}$ ${\cal O}{(1)}$ corresponds on the field theory side to a baryon density of the order ${\cal O}{(\lambda^{2})}$. This is the precise meaning of the parametrically large baryon densities mentioned earlier in the text.

In the deconfined phase it is the bulk black hole that sources the flux, but at zero temperature we need to introduce charges (the holographic baryons) that will source it. Following Witten's picture of the holographic baryons \cite{Witten:1998xy}, the charges are $N_c$ end points of strings stretched between the flavor D8-brane and a D4-brane wrapping the $S^{4}$ part of the geometry. The number $N_c$ follows from Gauss's law on $S^{4}$, including the flux generated by the stack of $N_c$ D4-branes, which give rise to the adjoint sector. 

The tension of the D4-D8 strings will cause the baryonic D4-branes to reside within the world volume of the D8-brane \cite{Seki:2008mu}. Hence, from the point of view of the holographic and field theory directions, the baryonic D4-branes appear as point-like particles carrying $N_c$ units of charge minimally coupled to the U(1) gauge field,
\be
N_c \int A_{a} \frac{d x^{a}}{d \tau} d \tau.
\ee
In order to obtain the mass, consider  the baryon world line with $t$ being the parameter on the geodesic. The action for it is identified with the DBI action for a D4-brane. There is a subtle issue here, namely that these D4-branes are wrapping $S^{4}$, whose size depends on the radial position. This leads to a mass that is a nontrivial function of the radial coordinate. The DBI action for the D4-brane is 
\be
S_{D4} = -\f{V_4 R^\f{15}{4} u^\f{1}{4}}{ (2 \pi)^4 g_s l_s^5}     \int dt  \sqrt{-g_{tt}},
\ee
where we are keeping $\sqrt{-g_{tt}}$ under the integral in order to compare with the general action for a point-like particle. After taking into account the fact that the action for a point-like massive particle at rest (i.e., $u^{t} = \f{1}{\sqrt{-g_{tt}}}$) with the world line parametrized by $t$ is
\be
S = -\int dt \left( m \sqrt{-g_{tt}} \right),
\ee
the mass and the charge of the holographic baryon read
\be
\label{eq.massandcharge}
m= \f{R^\f{15}{4} u(t)^\f{1}{4}}{ 6 \pi^2 g_s l_s^5}, \qquad q= N_c.
\ee
Note that in order to interpret $m$ from Eq. \eqref{eq.massandcharge} with $u(t) = u_{KK}$ as the mass of the baryon on the field theory side for the antipodal embedding of the D8-branes, one needs to take into the account the redshift factor $\sqrt{-g_{tt}}$ evaluated at $u = u_{KK}$. This leads to an expression for the baryon mass matching the ones in Refs. \cite{Rozali:2007rx} and \cite{Hata:2007mb} (the latter in the leading order in the large $\lambda$ expansion). The discussion generalizes trivially to nonantipodal embeddings.

\section{Motivations for going to parametrically large densities \label{sec.motivationforlargedensity}}

In this section we provide the rationale for studying holographic QCD at parametrically large baryon densities, i.e., of order of ${\cal O}{(\lambda^{2})}$. It turns out that there are several reasons for doing so.

In the first place, as was explained in the previous section, ${\cal O}{(\lambda^{2})}$ baryon densities are dual to a $O(\lambda)$ gauge field on the flavor brane. Such parametrically large gauge fields are the biggest ones that can be achieved on the flavor brane in the DBI approximation and as such require taking into account the nonlinear nature of the DBI action. This means, in particular, that the equations of motion for mesons (and also for \emph{any} other perturbations) will be modified, possibly leading to a different in-medium meson spectrum. Such an effect was preliminarily studied in Ref. \cite{Kim:2007zm} with an explicit simplifying assumption about the charge distribution on the flavor brane, namely that all the charge will be localized at $\tilde{z} = 0$. We will reconsider this effect in Sec. \ref{sec.mesons}, taking into account dynamical charge distributions.

It is also easy to notice, by examining the action \eqref{eq.SDBI} + \eqref{eq.SCS}, that only for ${\cal O}{(\lambda^{2})}$ baryon densities does the Chern-Simons term contribute to the equations of motion on an equal footing with the kinetic term. This has profound consequences for the low-energy physics of cold and dense holographic QCD, as the Chern-Simons term in the presence of an electric field is known to lead to instabilities towards creating striped phases \cite{Nakamura:2009tf,Ooguri:2010kt,Ooguri:2010xs,Bayona:2011ab}. Such instabilities modify the background and lead to different properties of the excitations. Although the operators that are condensing are different from those that condense in the quarkyonic chiral spirals \cite{Kojo:2009ha}, it is conceivable that the latter also condense in the expected way, but this effect might not be computable using the observables captured by the DBI + Chern-Simons action alone. Some further discussion
of this point is presented in the Conclusions. We will investigate the existence of instabilities by examining the presence of the marginally stable modes in Sec. \ref{sec.stability}.

On a related note, the Chern-Simons term leads to a particular mixing pattern between vector and axial mesons at nonzero momentum in the presence of background baryon density \cite{Domokos:2007kt}. Again, this can only happen for ${\cal O}{(\lambda^{2})}$ densities. We will return to this phenomenon in the Conclusions.

The most striking feature of our model at ${\cal O}{(\lambda^{2})}$ baryon densities, however, is the way the holographic baryons organize themselves on the flavor brane. Instead of forming a three-dimensional structure residing at $\tilde{z} = 0$, a very strong repulsion between the holographic baryons leads to a four-dimensional volume on the flavor brane, which is occupied by the holographic baryons. This was noticed already in Ref. \cite{Rozali:2007rx}, where the authors constructed a holographic background at ${\cal O}{(\lambda^{2})}$ baryon densities postulating a particular mean-field description. We will reintroduce this model in the next section and in the remaining part of this section we will focus on the implications of having parametrically large baryon densities, which follow from more microscopic considerations.

The elements of this microscopic understanding came about very recently in Ref. \cite{Kaplunovsky:2012gb}, where the authors considered simplistic models of baryonic lattices and preliminarily showed that for the same baryon density a lattice extending in the holographic direction has a lower free energy than lattices residing at the bottom of the flavor brane. The result of these studies are charge distributions formed out of a very few layers residing on top of each other in the holographic direction. The work of Ref. \cite{Rozali:2007rx} needs to be understood as a postulated continuum model of the situation in which very many such layers are formed.

The crucial observation behind our work is that for ${\cal O}{(\lambda^{2})}$ baryon densities the interactions of mesons with the baryonic medium are enhanced to the level that they influence the spectrum of mesons. Although a single baryon's response to the flavor-brane gauge-field perturbation is \mbox{$\lambda$-suppressed} due to the factor of $\lambda$ sitting in its mass, large number of baryons residing in a fixed holographic volume makes the overall contribution an ${\cal O}{(1)}$ effect, and \emph{a priori} equally as important as the kinetic term for the gauge-field perturbations. We will describe this phenomenon in detail in Sec. \ref{sec.mesons}.

We want to stress at this point that a fully reliable macroscopic description of ${\cal O}{(\lambda^{2})}$ baryon densities does not currently exist. Because of this, we will reuse the coarse-grained description of Ref. \cite{Rozali:2007rx}. In the next section we review the idea behind it and point out possible caveats.

\section{An attempt towards constructing mean-field description of dense hQCD\label{sec.meanfieldattempt}}
\subsection{Mean-field Lagrangian\label{sec.meanfielddesc}}
Having introduced the holographic baryons into the model and having understood the motivation for studying parametrically large ${\cal O}{(\lambda^{2})}$ baryon densities, we can proceed to a coarse-grained description of the bulk physics.

A single baryon sits at the bottom of the D8-brane, as this minimizes its gravitational potential energy coming from the D8-brane being nontrivially curved. As more baryons are added, initially they all stay at the bottom of the D8-brane and get distributed only along the field theory directions\footnote{As in the $N_f = 1$ Sakai-Sugimoto model baryons repel each other, one should think about the dual field theory at finite baryon density as being defined in a finite volume, e.g. on a torus.}, as climbing the holographic direction costs $\lambda$-enhanced gravitational potential energy. However, for sufficiently large densities, the strong repulsion between the holographic baryons will start organizing them not in three-, but rather four-dimensional structures \cite{Rozali:2007rx,Kaplunovsky:2012gb}. 

This can be understood in simple terms by noting that the baryons residing at the bottom of the D8-brane generate an electric field with a nonzero radial component oriented towards the boundary. Imagine now adding a single baryon on top of the others already residing at $\tilde{z} = 0$ and distributed in the field theory directions. If the density is large enough, then the electric repulsion between the $\tilde{z}=0$ baryons and a new baryon that we are trying to add to the system is such that the equilibrium position for the added baryon is no longer at $\tilde{z} = 0$, but rather at some $|\tilde{z}|>0$. As the baryon mass scales as $N_{c} \, \lambda$ and its charge as $N_{c}$, representing a nonzero baryon density in the dual field theory in terms of the three-dimensional density of the holographic baryons residing at $\tilde{z} = 0$--which is the description adopted in the references \cite{Bergman:2007wp,Kim:2007zm}--will be the relevant way of describing the system only for a ${\cal O}{(1)}$ baryon density in the dual field theory. 

For parametrically larger values of the baryon density, the bulk description in terms of the four-dimensional density of the holographic baryons will be appropriate. Such a way of describing the system was adopted for the first time for antipodal D8-brane embedding in Ref. \cite{Rozali:2007rx}, and here we will generalize this picture to the nonantipodal case.

In Ref. \cite{Rozali:2007rx} it was also conjectured that the holographic baryon density--being not three- but rather four-dimensional--is a dual manifestation of building up the quark Fermi surface or even the direct gravity dual of the Fermi surface. Indeed, the bulk charge distribution climbing up in the bulk radial direction, which is associated with the energy scale in the dual field theory, makes the connection with building up the Fermi surface tempting. There is, however, another motivation for this association. Although the baryon size (on the boundary) is ${\cal O}{(1)}$ in both $N_{c}$ and $\lambda$, and hence baryons can significantly overlap at relatively low, three-dimensional bulk densities, for such configurations the holographic system is not in a strongly correlated phase. The latter feature--strong interactions between the holographic baryons--arises precisely at ${\cal O}{(\lambda^{2})}$ densities, which is in line with the point of view presented in Ref.
\cite{Kaplunovsky:2012gb}.

The holographic system corresponding to parametrically large baryon densities, after the Kaluza-Klein reduction on the four-sphere, is a set of large numbers of charged point-like particles interacting via the U(1) gauge field with the DBI kinetic term. We are interested in the coarse-grained description of this system, as superficially only long-wavelength features will be important for understanding instabilities associated with the presence of the Chern-Simons term and nonzero density, as well as the meson perturbation, whose frequencies in the vacuum are O(1) in both $N_{c}$ and $\lambda$. 

There are general arguments following from large-$N_{c}$ counting that the ground state structure of such a system is crystalline \cite{Kaplunovsky:2012gb}. These arguments are based on the kinetic energy being suppressed by the $N_{c}$-heavy mass of a single (holographic) baryon while the potential energy is enhanced by the $N_{c}$-large quark charge. Below we will review the analysis of Ref. \cite{Rozali:2007rx}, where, setting aside certain subtleties that we will discuss, the macroscopic baryon densities--four-dimensional holographic crystals--were constructed in a simple coarse-grained description. The key idea there was to minimize the total energy of the dynamical electromagnetic field generated by the continuous charge distribution, as well as the potential energy of the charges in the dynamical electromagnetic and external gravitational fields. Such a description bears similarity to the AdS/CMT electron star construction \cite{Hartnoll:2010gu} and in some respects can be understood as a probe-brane analogue of it.

In the coarse-grained description, the relevant degrees of freedom will be the gauge field (as it encodes the axial and the vector currents), as well as the quantities describing a continuous charge distribution, i.e., the local energy density $\epsilon$ and local charge density $\rho$. Note that the gauge field we will be talking about will be a mean-field one. Nevertheless, we will explicitly assume that its kinetic term is still given by the DBI Lagrangian, which is the case in the vacuum.

As the system is put in an external gravitational field (the curved D8-brane world volume), we will take the redshift into account by introducing the local velocity of the charge density $u^{a}$, which is normalized as
\be
\label{eq.unorm}
u^{a} u_{a} = -1.
\ee
Although $u^{0}$ encodes the redshift factor, which is of order ${\cal O}{(1)}$ in $N_{c}$ and $\lambda$, it is important to note that we will never be interested in having ${\cal O}{(1)}$ components of the velocity field in the radial direction, as well as in the spacelike field theory directions (up to boosts, which are trivial). The reason for this is that the crystal-like nature of the configurations we study would not allow such motion to occur. However, infinitesimal--of order $O(1/\lambda)$--displacements in spacelike directions are not only permitted, but also lead to new interesting physical effects, which we will preliminarily study in Sec. \ref{sec.mesons}.

The missing ingredient in our description is the equation of state that relates the energy and charge density. For a free system both quantities are proportional to the number density, with the proportionality constants being respectively the mass and the charge of a single particle. In the interacting system, the energy density can be corrected by interactions between the constituents in a given infinitesimal volume. Another way of looking at this is that by using the coarse-grained gauge field in the mean-field approach we are no longer including in the total energy count of the system the highly oscillatory behavior of the microscopic gauge field coming from the granular structure of the substance under consideration. In order to better investigate this issue, let us assume  that the local energy density $\epsilon$ and the local charge density $\rho$ are proportional to the local baryon density $n$,
\be
\label{eq.freeEOS}
\epsilon = m_{b}(z) \, n \quad \mathrm{and} \quad \rho = q_{b} \, n,
\ee
and estimate the mistake made by neglecting interactions within the small volume element $\delta V$.

Note that only for a holographic number density of order ${\cal O}{(\lambda^{2})}$ do the energy density, the potential energy of charges proportional to $\rho A_{0}$, and the kinetic term for the gauge field all contribute on an equal footing. The ${\cal O}{(\lambda^{2})}$ density of the holographic baryons results in a holographic lattice spacing $d$ of order $O(\lambda^{-1/2})$. We now wish to examine whether the mean-field description is a reasonable approximation. To do so,
we should compare the interactions of a collection of lattice points in a small volume $\delta V$ to those of a smeared charge distribution. 

The interactions of a collection of lattice points can be estimated by (i) assuming that the interactions are not those of a DBI theory, but rather of five-dimensional electromagnetism, and
(ii) assuming that we are in flat space. An approximate expression is thus
\be
\label{eq.deltaeps}
\delta \epsilon \approx n \sum_{\delta V} \frac{N_{c}}{\lambda} \frac{1}{d^{2} \left( p_{1}^{2} + p_{2}^{2} + p_{3}^{2} + p_{4}^{2} \right)},
\ee
where the $p_{j}$'s denote integer node numbers in the $\tilde{z}$ and $\tilde{x}^{i}$ directions and we sum over all the holographic baryons residing in the volume $\delta V$. The sum over all nodes in Eq. \eqref{eq.deltaeps} diverges as the square root of the total number of nodes $\#_{nodes}$
\be
\label{eq.deltaepsnodes}
\delta \epsilon \approx N_{c} n \sqrt{\#_{nodes}},
\ee
or, after using $\#_{nodes} \approx n \, \delta V$, as
\be
\label{eq.deltaepsdeltaV}
\delta \epsilon \approx N_{c} n^{3/2} {\delta V}^{1/2}.
\ee
If we replace the lattice by a continuum distribution, we will get an answer that scales in a similar way, as will the difference between the two. 
There are two ways of looking at Eqs. \eqref{eq.deltaepsnodes} and \eqref{eq.deltaepsdeltaV}. As $n$ is ${\cal O}{(\lambda^{2})}$, for any finite $\delta V$ the changes in the energy density given by Eq. \eqref{eq.deltaepsdeltaV} are comparable to the energy density we started with, i.e., ${\cal O}(N_{c} \lambda^{3})$. Hence, if we are supposed to evaluate Eq. \eqref{eq.deltaepsdeltaV} for some finite/fixed volume $\delta V$, then the equation of state will receive order-one corrections and the mean-field approximation should not be used. On the other hand--and this is the approach we take in the remaining part of the text--if we are supposed to set the volume to be infinitesimally small, then the difference in energies \eqref{eq.deltaepsdeltaV} becomes negligible. An alternate way of phrasing this is that for a large but fixed number of nodes the energy in Eq.\eqref{eq.deltaepsnodes} will always be coupling-constant suppressed with respect to the energy density of the free system. Note finally that the volume-divergent contribution of the kind \eqref{eq.deltaeps} would be present for arbitrarily weak electromagnetic interactions; however, the weakness of the interactions then guarantees that evaluating it for a $\delta V$ of order ${\cal O}{(1)}$ still leads to a negligible contribution. This is precisely the rationale adopted in the electron star construction \cite{Hartnoll:2010gu}.

Having discussed a subtlety related to the equation of state, we will stick with the equation of state \eqref{eq.freeEOS} and in the remaining part of this subsection we will discuss the action principle describing our system. As the holographic baryon density $n$ is by definition non-negative, it will be useful, in the following, to express it as
\be
n = w^2.
\ee
Keeping in mind that we will never be interested in velocities $u^{a}$ having finite components in the radial and spacelike field theory direction, the action principle that describes our system is that of a charged dust \cite{Brown:1992kc}
\bea
\label{eq.Sdust}
S_{dust} = && \int d^{5} x \sqrt{-\mathrm{det}(g_{a b})} \big\{-m_{b}(z) w^2 + \nonumber \\ && + q_{b} w^2 u^{a} \left( A_{a} - \partial_{a} \phi \right) + \lambda (u_{a} u^{a} + 1) \big\},
\eea
where $q_{b} w^2 u^{a} A_{a} $ is the standard coupling of the charge current $q_{b}\, w^2 \, u^{a}$ to the $U(1)$ gauge field $A_{a}$; $\phi$ transforms as a phase under $U(1)$ gauge rotations and ensures on-shell conservation of the current, whereas the introduction of $\lambda$ takes care of the on-shell velocity  normalization \eqref{eq.unorm}. After introducing rescaled variables \eqref{eq.rescalings1}, the dust action becomes 
\bea
\label{eq.Sdustresc}
S_{dust}/c = && \int d^{5} \tilde{x} \sqrt{-\mathrm{det}(\tilde{g}_{a b})} \Big\{ -\beta \, \tilde{\xi}^{1/4} \big(1+\frac{\tilde{z}^2}{\tilde{\xi}^2}\big)^{1/12} \tilde{w}^2 \nonumber\\
&& + \gamma \, \tilde{w}^2 \tilde{u}^{a} \left( \tilde{A}_{a} - \partial_{a} \tilde{\phi} \right)+ \tilde{\lambda} (\tilde{u}_{a} \tilde{u}^{a} + 1) \Big\},
\eea
where
\be
\label{eq.betagammawrescalings}
\beta = 4\pi^2, \quad \gamma = 12\pi^2 \quad \mathrm{and} \quad \tilde{w} = \frac{2\pi l_{s}^{2}}{R^{2}} w.
\ee
The rescaling of other variables will not be important for our results, but it can be easily worked out. Note again that the rescalings \eqref{eq.betagammawrescalings} imply that the ${\cal O}(1)$ rescaled baryon number density $\tilde{w}^2$ corresponds to an order ${\cal O}{(\lambda^{2})}$ physical baryon density $w^2$. As anticipated earlier in the text, rescalings \eqref{eq.rescalings1} and \eqref{eq.betagammawrescalings} take us directly to the very dense phase with baryons climbing up the holographic direction. For rescaled densities that are $\lambda$-suppressed, the terms describing baryon dynamics drop out and baryons then reside at $\tilde{z} = 0$ and appear as boundary conditions for the flux, exactly as in Refs. \cite{Bergman:2007wp,Kim:2007zm}. We think, however, that in the latter cases the backreaction of the baryons on the embedding should be treated as a $\lambda$-suppressed effect. Otherwise, instead of taking it into account, one should rather pass to the effective description of four-dimensional holographic baryon densities, as is done here and originally in Ref. \cite{Rozali:2007rx}.

The full action is the sum of $S_{DBI}$ and $S_{dust}$ and its variation leads to the equations of motion
\begin{subequations}
\bea
\label{eq.eomsfordust4phi}
&&\frac{\delta S}{\delta \tilde{\phi}}:\,\nabla_{a} (\gamma \tilde{w}^2 \tilde{u}^{a}) = 0, \\
\label{eq.eomsfordust4w}
&&\frac{\delta S}{\delta \tilde{w}}:\,\beta \, \tilde{\xi}^{1/4} \big(1+\frac{\tilde{z}^2}{\tilde{\xi}^2}\big)^{1/12} \tilde{w} = \gamma \, \tilde{w} \tilde{u}^{a} ( \tilde{A}_{a} - \partial_{a} \tilde{\phi}),\quad \\
\label{eq.eomsfordust4lambda}
&&\frac{\delta S}{\delta \tilde{\lambda}}:\,\tilde{u}_{a}\tilde{u}^{a} = -1, \\
\label{eq.eomsfordust4ua}
&&\frac{\delta S}{\delta \tilde{u^{a}}}:\,\gamma \tilde{w}^2 \left( \tilde{A}_{a} - \partial_{a} \tilde{\phi} \right) + 2 \tilde{\lambda} \tilde{u}_{a} = 0.
\eea
\end{subequations}
These equations need to be supplemented with the DBI generalization of Maxwell's equations for $\tilde{A}_{a}$,
\bea
\label{eq.DBIgeneralizationofMaxwell}
&&\frac{\partial}{\partial \tilde{x}^{a}}\frac{\partial}{\partial(\partial_{a} \tilde{A}_{b})} \left(\tilde{\xi}^{1/4} \big(1+\frac{\tilde{z}^2}{\tilde{\xi}^2}\big)^{1/12} \sqrt{-\mathrm{det}(\tilde{g}_{c d} + \tilde{F}_{c d})}\right) = \nonumber\\
&& = - \sqrt{-\mathrm{det}(\tilde{g}_{c d})} \gamma \tilde{w}^2 \tilde{u}^{b} - 3 \, \alpha \, \epsilon_{a c d e f} \tilde{F}^{c d} \tilde{F}^{e f},
\eea
which comes from varying the full action with respect to $\tilde{A}_{a}$, and from the equation of motion for $\tilde{y}$ entering the action through the induced metric $\tilde{g}_{a b}$. 

Note that Eqs. \eqref{eq.eomsfordust4ua} and \eqref{eq.DBIgeneralizationofMaxwell} imply that there is an explicit coupling between the gauge field--in particular the gauge field perturbations corresponding to the mesons--and the holographic baryon density. The subtlety that the velocities in the spacelike directions cannot be taken finite due to the underlying crystal structure is explained in Sec. \ref{sec.mesons}. There might also be gradients that we have neglected at this point, and we elaborate on their role for meson perturbations in Sec. \ref{sec.mesons}. 

Setting aside those subtleties, in the next subsection we will use the action \eqref{eq.Sdustresc} to construct homogeneous large-baryon-density states for the family of models parametrized by the asymptotic separation $L$ of the D8- and $\overline{\mathrm{D8}}$-brane. We will show that the resulting background (in the antipodal case) is exactly the one studied in Ref. \cite{Rozali:2007rx}.

\subsection{Solutions for embeddings\label{subsec.solnsforembeddings}}
In equilibrium, the most natural assumption for the long-wavelength description of a nonzero baryon density state is that of isotropy and homogeneity\footnote{We will try to relax these in the next section when looking for marginally stable inhomogeneous modes.}. This means that none of the functions appearing in the action \eqref{eq.Seff} + \eqref{eq.Sdustresc} depend on the field theory directions $\tilde{x}^{\mu}$. Moreover, in the rest frame the symmetries dictate that the only nontrivial component of the gauge field is $\tilde{A}_{0}$, and the same is true for the baryon velocity, i.e., $\tilde{u}^{0}$. 

The velocity needs to be subject to the normalization condition \eqref{eq.unorm}, which in this situation fixes its form completely,
\be
\tilde{u}^{0} = \frac{1}{\tilde{\xi}^{1/4} (\tilde{\xi}^2 + \tilde{z}^2)^{1/4}}.
\ee
Substituting this result into the action \eqref{eq.Sdust} and setting $\tilde{\phi}$ to 0 results in an effective action for the embedding function $\tilde{y}$, the baryon number density $\tilde{w}^2$, and the gauge field $\tilde{A}_{0}$ of the form
\bea
\label{eq.Seffreduced}
&&\tilde{S}_{eff} / c = \int d\tilde{z} d^{4} \tilde{x} \Big\{ - \tilde{\xi}^{1/4} \big(1+\frac{\tilde{z}^2}{\tilde{\xi}^2}\big)^{1/12} \times \nonumber\\
&& \sqrt{-\mathrm{det}(\tilde{g}_{ab}) - \tilde{\xi}^{9/2}  \big(1+\frac{\tilde{z}^2}{\tilde{\xi}^2}\big)^{3/2} \tilde{A}_{0}'(\tilde{z})^{2}} - \nonumber\\&& \beta \sqrt{-\mathrm{det}(\tilde{g}_{ab})} \, \tilde{\xi}^{1/4} \big(1+\frac{\tilde{z}^2}{\tilde{\xi}^2}\big)^{1/12} \, \tilde{w}(\tilde{z})^2 + \nonumber\\ 
&&\gamma \sqrt{-\mathrm{det}(\tilde{g}_{ab})} \, \tilde{\xi}^{-1/4} \big(1+\frac{\tilde{z}^2}{\tilde{\xi}^2}\big)^{-1/4}\tilde{w}(\tilde{z})^2 \tilde{A}_{0}(\tilde{z})
 \Big\},
\eea
where $\mathrm{det}(\tilde{g}_{ab})$ is the determinant of the induced metric \eqref{eq.inducedmetric}. The action \eqref{eq.Seffreduced}, in the antipodal case [i.e., $\tilde{y}'(\tilde{z}) = 0$], is identical to the one used in Ref. \cite{Rozali:2007rx} after suitable redefinitions of what is meant by the mass of the holographic baryons and their density\footnote{As opposed to Ref. \cite{Rozali:2007rx}, in our approach the mass of the baryonic D4-brane comes without the factor of $\sqrt{-\tilde{g}_{00}}$, and $\tilde{u}^{0} = 1/\sqrt{-\tilde{g}_{00}}$ is not absorbed in what is meant by the density ($\tilde{w}^2$). Moreover, in our effective Lagrangian the electromagnetic coupling comes with $\tilde{u}^{0} = 1/\sqrt{-\tilde{g}_{00}}$ factored out.}. 

We are interested in studying the theory at nonzero baryon density and vanishing axial density. We thus focus on the symmetric (with respect to $\tilde{z} \leftrightarrow - \tilde{z}$) configurations of $\tilde{A}_{0}(\tilde{z})$, which, through flux conservation, need to be supported by a (symmetric) profile of $\tilde{w}(\tilde{z})$.

The curvature of the D8-brane will make the holographic baryons fall towards $\tilde{z} = 0$, and the U(1) interactions between them will counter this effect. So the expectation is that the holographic baryon density will reside at the bottom of the U-shaped D8-brane. In the following, we will derive the properties of the holographic charge distribution from the equations of motion for the effective action \eqref{eq.Seff}. A technical improvement, compared to Ref. \cite{Rozali:2007rx}\footnote{Apart from considering nonantipodal D8-brane embeddings.}, is the use of the positivity of the number density [$\tilde{n}(\tilde{z}) = \tilde{w}(\tilde{z})^2$] to \emph{explicitly} extremize the action. We will do this by solving the equations of motion separately inside and outside the holographic baryon density and matching the solutions.

Indeed, let us start by considering the equation of motion obtained by taking the functional derivative of Eq. \eqref{eq.Seff} with respect to $\tilde{w}(\tilde{z})$, which after trivial simplifications reads 
\be
\label{eq.dSeffdw}
\tilde{w}(\tilde{z}) \left\{\gamma \, \tilde{A}_{0}'(\tilde{z}) - \beta \, \tilde{\xi} \big(1+\frac{\tilde{z}^2}{\tilde{\xi}^2}\big)^{1/3} \right\} = 0.
\ee
If $\tilde{w}(\tilde{z}) = 0$, this equation is trivially satisfied. However, for $\tilde{w} > 0$ (i.e., in the presence of the macroscopic density of the holographic baryons) Eq. \eqref{eq.dSeffdw} fixes the form of $\tilde{A}_{0}(\tilde{z})$,
\be
\label{eq.A0in}
\tilde{A}_{0}(\tilde{z}) = \frac{\beta}{\gamma} \, \tilde{\xi} \big(1+\frac{\tilde{z}^2}{\tilde{\xi}^2}\big)^{1/3}.
\ee
The ratio of $\beta$ to $\gamma$ is $1/3$, but in Eq. \eqref{eq.A0in} we will keep $\beta$ and $\gamma$ as explicit parameters. Our motivation for doing this
is that it is conceivable that a more precise microscopic description of the system will lead to an effective mean-field description not dissimilar from
the one we have been employing, but with different values for $\beta$ and $\gamma$, and as we will see this may lead to important qualitative changes
in the behavior of the system. 

Consider now the equation of motion for $\tilde{y}(\tilde{z})$. This quantity appears in the effective action \eqref{eq.Seff} only through the determinant of the metric. Note also that the matter part of Eq. \eqref{eq.Seff} vanishes on-shell by Eq. \eqref{eq.dSeffdw}. This implies that the embedding function $\tilde{y}(\tilde{z})$ knows about the matter on the D8-brane only through the electric field sourced by it. 

The equation of motion for $\tilde{y}(\tilde{z})$ again has a first integral, which reads
\bea
\label{eq.yfirstintegrat}
\frac{\tilde{\xi} ^{9/4} \big(1+\frac{\tilde{z}^2}{\tilde{\xi}^2}\big)^{19/12} \left(\tilde{\xi}^3+\tilde{\xi} \tilde{z}^2-1\right) \tilde{y}'(\tilde{z})}
{\sqrt{-\mathrm{det}(\tilde{g}_{ab})-\tilde{\xi}^{9/2}\big( 1+\frac{\tilde{z}^2}{\tilde{\xi}^2} \big)^{3/2} \tilde{A}_{0}'(\tilde{z})^{2}}} = \sqrt{\tilde{\xi}^3 - 1}. \quad \quad 
\eea
The rhs is determined from the power series expansion of the lhs around $\tilde{z} = 0$ using Eq. \eqref{eq.A0in}. Equation \eqref{eq.yfirstintegrat} allows us to algebraically solve for $\tilde{y}'(\tilde{z})$ in terms of $\tilde{\xi}$ and $\tilde{A}_{0}'(\tilde{z})$. 

The only remaining equation of motion is the one for $\tilde{A}_{0}(\tilde{z})$, which reads
\bea
\label{eq.dSeffdA0}
&&\frac{d}{d\tilde{z}}\frac{\partial}{\partial \tilde{A}_{0}'(\tilde{z})} \Bigg\{ - \tilde{\xi}^{1/4} \big(1+\frac{\tilde{z}^2}{\tilde{\xi}^2}\big)^{1/12} \times \nonumber\\
&&\sqrt{-\mathrm{det}(\tilde{g}_{ab}) - \tilde{\xi}^{9/2}  \big(1+\frac{\tilde{z}^2}{\tilde{\xi}^2}\big)^{3/2} \tilde{A}_{0}'(\tilde{z})^{2}} \Bigg\} = \nonumber \\ &&\gamma \sqrt{-\mathrm{det}(\tilde{g}_{ab})} \, \tilde{\xi}^{-1/4} \big(1+\frac{\tilde{z}^2}{\tilde{\xi}^2}\big)^{-1/4}\tilde{w}(\tilde{z})^2.
\eea
After evaluating both sides of Eq. \eqref{eq.dSeffdA0} we use Eq. \eqref{eq.yfirstintegrat} to eliminate the dependence on the embedding function $\tilde{y}(\tilde{z})$. Inside the volume occupied by holographic charges Eq. \eqref{eq.dSeffdA0} specifies the charge density, as the gauge field is given by \eqref{eq.A0in}. On the outside, the rhs is zero and the first integral of the equations is the dimensionless flux $\tilde{E}$ through one end of the D8-brane,
\bea
\label{eq.A0out}
&&\frac{\partial}{\partial \tilde{A}_{0}'(\tilde{z})} \Bigg\{ - \tilde{\xi}^{1/4} \big(1+\frac{\tilde{z}^2}{\tilde{\xi}^2}\big)^{1/12} \times \nonumber\\
&&\sqrt{-\mathrm{det}(\tilde{g}_{ab}) - \tilde{\xi}^{9/2}  \big(1+\frac{\tilde{z}^2}{\tilde{\xi}^2}\big)^{3/2} \tilde{A}_{0}'(\tilde{z})^{2}} \Bigg\} = \tilde{E},\nonumber
\eea
or, equivalently, the baryon density $B$ in the dual field theory \eqref{eq.Bdef},
\be
B = \frac{\lambda^2}{96 \pi^4 M_{KK}^2 R^{5}} \tilde{E}.
\ee

The full solution is now obtained by matching the electric field, i.e., $\tilde{A}_{0}'(\tilde{z})$, inside the charge distribution on the D8-brane with the electric field of constant flux outside it. The UV parameters are $L$--the asymptotic separation of the D8- and $\overline{\mathrm{D8}}$-brane--specifying the dual theory and $\tilde{E}$, which specifies the state. The chemical potential in this approach is a secondary object, as it arises from the integral of $\tilde A_{0}'(z)$ from $\tilde{z} = 0$ to $\tilde{z} = \infty$ with the boundary condition that $\tilde{A}_{0}(0)$ is given by the formula \eqref{eq.A0in} evaluated at $\tilde{z} = 0$. There are two IR parameters in the procedure: $\tilde{z}_{0}$--the radial position on the D8-brane where the charge density terminates--and $\tilde{\xi}$--the radial position in the target space of the lowest point of the D8-brane.

The solutions parametrized by $L$ and $\tilde{\rho}$ (or equivalently the chemical potential $\tilde{\mu}$) are obtained numerically by specifying $\tilde{\rho}$ and searching for $\tilde{\xi}$, such that the formula \eqref{eq.Lintermsofy} with $\tilde{y}'(\tilde{z})$ given by Eq. \eqref{eq.yfirstintegrat} is satisfied. Note that for each checked $\tilde{\xi}$ one still needs to match the electric field at the boundary of the holographic charge distribution.

\begin{figure}
\includegraphics[width=8.5cm]{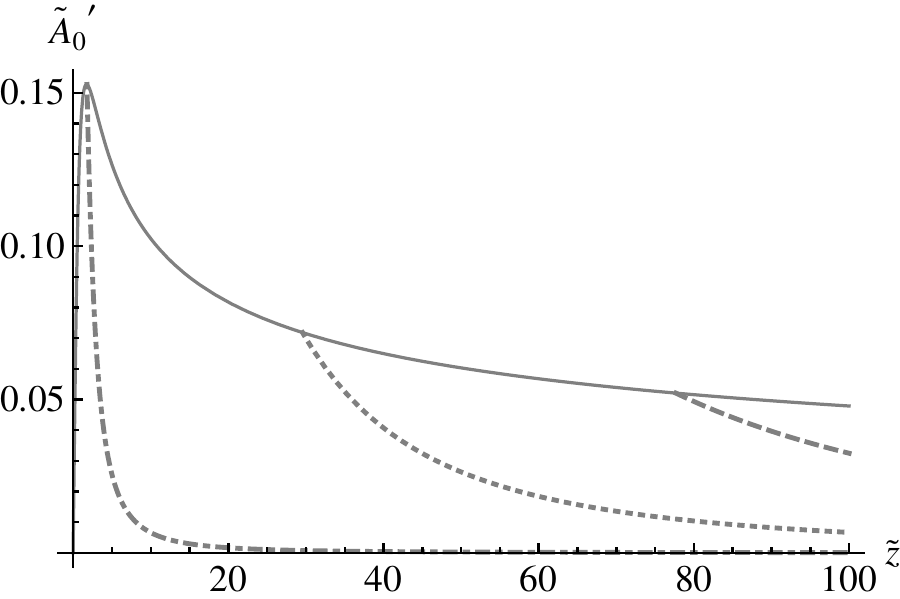}
\caption{The electric field $\tilde{A}_{0}'(\tilde{z})$ as a function of the radial coordinate $\tilde{z}$ in the antipodal case ($\tilde{\xi} = 1$) for three different rescaled baryon densities: $\tilde{E} = 1$ (dotdashed), $\tilde{E} = 100$ (dotted), and $\tilde{E} = 500$ (dashed). The envelope (solid curve) is the electric field inside the charge distribution. The discontinuity in the derivative corresponds to the radial position at which the holographic charge density terminates. The electric field never gets large, as the  matching condition bounds it from above. Note that the volume occupied by holographic charges increases with the baryon density on the field theory side, but the distribution of the charge in the core is not altered by the outer layers, which is why we used the envelope.}
\label{fig.electricfieldasafunctionofrho}
\end{figure}

\begin{figure}
\includegraphics[width=8.5cm]{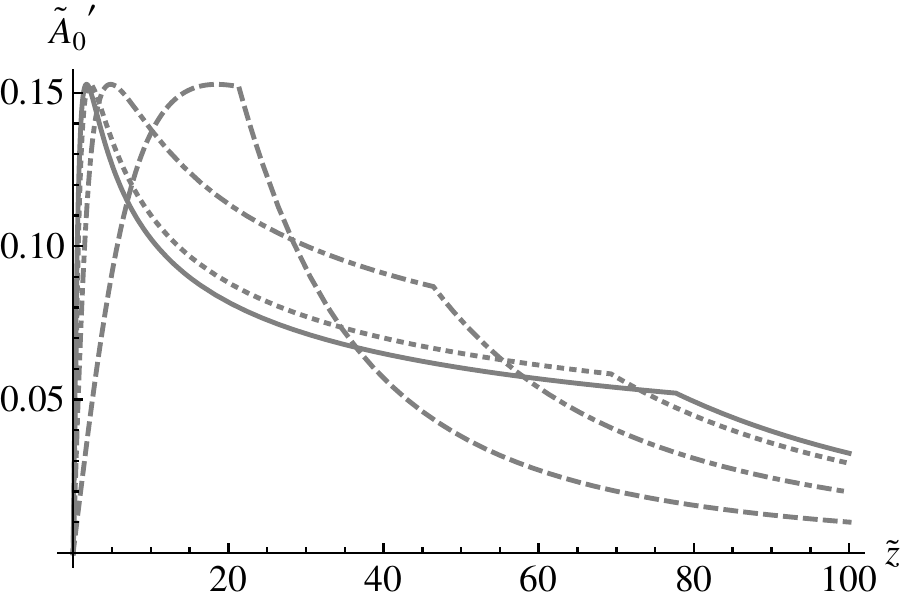}
\caption{The electric field $\tilde{A}_{0}'(\tilde{z})$ as a function of the radial coordinate $\tilde{z}$ at fixed baryon density ($\tilde{E} = 500$) for four different embeddings, ranging from the antipodal ($L/\delta\tau = 0.5$) to very nonantipodal ($L/\delta\tau = 0.05$): $L/\delta\tau = 0.5$ (solid), $L/\delta\tau = 0.2$ (dotted), $L/\delta\tau = 0.1$ (dotdashed), and $L/\delta\tau = 0.05$ (dashed). As in Fig. \ref{fig.electricfieldasafunctionofrho}, a discontinuity in the derivative corresponds to the radial position at which the holographic charge density terminates. The electric field also never gets large in this case also due to the matching condition at the boundary of the bulk charge distribution.}
\label{fig.electricfieldasafunctionofxi}
\end{figure}

Figure \ref{fig.electricfieldasafunctionofrho} shows the profile of the electric field for the antipodal embedding for four different charge densities, whereas Fig. \ref{fig.electricfieldasafunctionofxi} depicts the same quantity for $\tilde{E} = 500$ and four different embeddings from antipodal to very nonantipodal. What one can easily notice is that the electric field is bounded from above and never gets large enough to successfully compete with the induced metric in the DBI action. This can be easily understood, as the electric field is continuously matched from the electric fields inside and outside the charge distribution. In such a case, the electric field is bounded by the (smaller of) the maximum value(s) of the electric field inside or outside the charge distribution. By looking at Figs. \ref{fig.electricfieldasafunctionofrho} and \ref{fig.electricfieldasafunctionofxi} we see that the relevant number is about $0.15$, corresponding to the point where the electric field inside the charge distribution has a maximum with a vanishing first derivative. Indeed, although the position at which the second derivative of Eq. \eqref{eq.A0in} vanishes depends on the embedding through $\xi$, the value of the electric field at the maximum is fixed and reads
\be
\label{eq.maxelectricfield}
\tilde{A}_{0}'(\tilde{z}) \leq \frac{\beta}{2^{1/3}\sqrt{3} \gamma} \approx 0.153.
\ee

As argued earlier, the embedding feels the charges on the D8-brane only through the DBI term, and hence it never gets significantly distorted from its vacuum form by the presence of baryon charges. The smallness of the electric field  also has profound consequences on the stability of the system, as will be discussed in the next section.

Although in the top-down model of Sakai and Sugimoto the values of $\beta$ and $\gamma$ are fixed and are given by Eq.\eqref{eq.betagammawrescalings}, one can also treat them in the bottom-up spirit as free parameters. In particular, such an approach allows us to obtain values of the electric field on the D8-brane larger than $0.153$, and at the same time obtain denser distributions of the holographic baryons. This has important consequences for the stability of the system, as well as the form of the meson spectrum at nonzero density. Note also that for completely arbitrary $\beta$ and $\gamma$ we are not guaranteed that the matching condition (i.e., demanding the continuity of the electric field) at the bulk boundary of the holographic charge distribution will be satisfied.

Before we conclude this section, let us comment on and stress some aspects of the physics described by the model considered.

In the first place, note that the gauge field inside the volume of charges depends mildly on the total amount of charge--the latter appears through the matching condition at the boundary of the charge distribution, $\tilde{z} = \tilde{z}_{0}$, constrained by the fixed asymptotic separation of the D8-brane. This, through Eq. \eqref{eq.dSeffdA0}, implies that the way the charge is distributed does not change significantly\footnote{For antipodal embedding it does not change at all; for nonantipodal ones it changes through the change of $\tilde{\xi}$ for each $B$.} with the increase of the baryon density $B$. Instead, the charge added to the system accumulates on top of the existing one without drastically changing the original distribution. Although naive intuition would suggest that the charge at the bottom gets more and more compressed as we add more of it to the system, it is not correct and does not take into account the fact that there are also charges on the $\tilde{z}<0$ part of the U-shaped D8-brane countering this effect.

Let us consider a simple toy model which illustrates this effect: a system of charged point particles in four spatial dimensions with $1/r^2$ interactions, and with a
harmonic oscillator potential in one of the four spacelike coordinates which we call $u$. One can easily show that in this toy model all finite-density configurations have exactly the same constant density and are only distinguished by their extent in the $u$ direction, similar to what we see in the Sakai-Sugimoto model. 

It is also very interesting and puzzling that in the bulk we are describing a finite-density system with fermionic constituents (quarks) in terms of geometric objects (D4-branes) that in the vacuum correspond to bound states of the fermions. Note that from the point of view of the dual field theory baryons are (heavily) overlapping; nevertheless, in the bulk this seems to be taken into account by a reasonably simple coarse-grained description. The caveat might be that such a description is simple because it misses some ingredients. For example, in the non-Abelian case with $N_{f}>1$ the holographic baryons are described by instantons on the stack of the flavor branes, and these presumably interact through complicated many-body interactions stemming from the nonlinear nature of the DBI action.

\section{Homogeneity of the dense hQCD at large distances \label{sec.stability}}

\subsection{The role of the Chern-Simons term}

So far in our description of cold baryonic matter in the Sakai-Sugimoto model we explicitly assumed homogeneity at large distances in both the charge distribution and the ansatz for the mean-field gauge field. These are usually sensible assumptions, as typically gradient terms increase the (free) energy. However, it was recently found that the presence of the Chern-Simons term in holographic Lagrangians for $U(1)$ gauge fields (like in our case) together with a large enough radial electric field may lead to the appearance of tachyonic modes of the gauge field carrying nonzero momentum \cite{Domokos:2007kt,Nakamura:2009tf}. The condensation of such modes, after including nonlinear effects, is supposed to give a new inhomogeneous and stable state (or at least one having lower free energy than the homogeneous state). This was indeed verified in particular examples in Refs. \cite{Ooguri:2010kt} and \cite{Ooguri:2010xs}.

The effect in question seems to be generic and was found within a class of bottom-up holographic models having as a solution the Reissner-Nordstr\"om black hole in AdS$_5$ \cite{Nakamura:2009tf}, and it was also shown to exist in the very same setting that we are considering--the Sakai-Sugimoto model, albeit in the deconfined phase at large enough baryon density \cite{Ooguri:2010xs}, as well as at zero temperature and big enough axial chemical potential \cite{Bayona:2011ab}. It also turned out that similar phenomena can be observed in four-dimensional holographic models that include pseudoscalar fields and parity-violating terms for the gauge field \cite{Donos:2011bh}. Subsequent developments in this active area of research include \cite{Donos:2011qt,Donos:2011ff,Donos:2012gg,Donos:2012wi,Bergman:2011rf,Jokela:2012vn}.

We are not going to perform comprehensive analysis of the linear stability of our effective model, as this has not been achieved yet in holographic systems that are less complex than ours. Instead, following earlier works on the role of the Chern-Simons term in the Sakai-Sugimoto model \cite{Ooguri:2010xs,Bayona:2011ab}, we will search for the presence of a particular normalizable mode, such that it carries nonzero momentum, has a zero frequency, and its polarization ensures decoupling from other perturbations at the linearized level. Such a mode is called a marginally stable one, as typically perturbations carrying momentum evolve or disperse, whereas this one stays constant in time. It is expected\footnote{We verified this in the Sakai-Sugimoto model put at nonzero axial chemical potential, and expect it to be correct in other cases too.} that in the direct neighborhood of parameters for which marginally stable modes appear there will be tachyonic modes. In that sense, searching for marginally stable modes is testing some aspects of the stability of the system towards creating structures that break translational invariance. 

Based on previous studies in Refs. \cite{Ooguri:2010xs,Bayona:2011ab}, we will search for the presence of the following marginally stable mode:
\be
\label{eq.InstPert}
\delta \tilde A_\mu d\tilde x^\mu = \tilde h(\tilde z) \left\{\cos (\tilde k \tilde x_3) d \tilde x^1 + \sin (\tilde k \tilde x_3) d \tilde x^2\right\},
\ee
which will be a small perturbation on top of the finite-density background derived in Sec. \ref{subsec.solnsforembeddings}. Such a normalizable mode turned out to be present in the Sakai-Sugimoto model in the deconfined phase for a large enough baryon density and in the confined phase for large enough axial density. In both situations it was necessary to consider parametrically large--of order ${\cal O}(\lambda^{2})$--densities, as expected on general grounds and discussed in Sec. \ref{sec.motivationforlargedensity}. It is worth stressing already at this point that it was not the large rescaled density of charge which was really necessary in both cases, but rather the large bulk electric field associated with it. Note however, that in our case the electric field is bounded from above by a rather small value \eqref{eq.maxelectricfield}. The second comment in place is that the ansatz invoked \eqref{eq.InstPert} leads to the appearance of a nonzero magnetic field, but it does not modify the electric field in any way. As the magnetic field couples only to the moving charge, the charge distribution is not going to be altered by the mode \eqref{eq.InstPert}.

The breakdown of translational invariance at nonzero baryon density due to the presence of the Chern-Simons term in the flavor-brane Lagrangian might not be an unwanted feature after all, as the quarkyonic phase is expected to be inhomogeneous too. Although the condensate there \cite{Kojo:2009ha}, which was given by
\be
\langle \bar \psi \exp(2 i \mu  x_3 \gamma_0 \gamma_3 ) \psi \rangle
\ee
is clearly different from the prospective one in our case,
\be
\langle \bar \psi \left[ \gamma_1 \cos (\tilde k \tilde x_3) + \gamma_2 \sin  (\tilde k \tilde x_3) \right] \psi \rangle,
\ee
it is not entirely inconceivable that the modulation in the vector or axial current triggers the one encountered in the quarkyonic phase. We will come back to this issue in the Conclusions. Last but not least, it is also worth stressing that any inhomogeneities at large distances will influence the spectrum of mesons, and as the spectrum is another feature we are investigating the presence of instabilities is of clear importance and interest for us. 

To clarify the mechanism for the instability, we focus on the antipodal case where the expressions are less cluttered. Searching for the marginally stable mode \eqref{eq.InstPert} in the nonantipodal case is a simple generalization of the antipodal situation, and hence we skip the technical details and just present the results. It is instructive to derive the reduced action for perturbations by Eq. \eqref{eq.InstPert} by expanding our coarse-grained action to the quadratic order in $\tilde{h}$. As expected on physical grounds, we find that the perturbations in question decouple, leading to the reduced one-dimensional Lagrangian density
\be
\label{eq.LagInstability}
\mathcal L = a(\tilde{z}) \tilde{h}'(\tilde{z})^2 + b(\tilde{z}) \tilde{h}( \tilde{z})^2 + 12 \alpha \, \tilde{k} \, \tilde{h}(\tilde{z})^2 \tilde{A}_0'(\tilde{z}),
\ee
where
\bea
\label{eq.aANDbFUNCS}
a(\tilde{z}) &=& - \f{3(1+\tilde{z}^2)}{2 \sqrt{4- 9 (1+\tilde{z}^2)^{1/3} \tilde{A}_0'(\tilde{z})^2}}, \nonumber \\
b(\tilde{z}) &=& -\f{\tilde{k}^2 \sqrt{4 - 9 (1+\tilde{z}^2)^{1/3} \tilde{A}_0'(\tilde{z})^2}}{6(1+\tilde{z}^2)^\f{1}{3}}
\eea
and $\tilde A_0(\tilde z)$ is the background gauge field. It is easy to see already at the level of Eq. \eqref{eq.LagInstability} that the Chern-Simons term--the one entering with $\alpha$ coefficient--can lead to tachyonic behavior at nonzero momenta: although $b(\tilde{z})$ (playing the role of the mass term) is negative or zero as it needs to be, the Chern-Simons-term contribution does not come with a definite sign and in certain circumstances can win over the standard mass term.

Up to this point, while deriving Eq. \eqref{eq.LagInstability}, we did not assume anything about the form of the background gauge field; it could come from the axial or baryon charge, or both. Let us now discuss a subtlety that has not been present in the axial case analyzed in Ref. \cite{Bayona:2011ab} and which complicates studies of the baryonic case here. It is easy to spot that $a(\tilde z)$  and $b(\tilde z)$ are even under parity, i.e., $\tilde{z} \rightarrow - \tilde{z}$. We see that if $\tilde A_0(\tilde z)$ is odd under parity, which corresponds to purely axial chemical potential, the Lagrangian \eqref{eq.LagInstability} is even under parity. This implies that the solutions $\tilde h(\tilde z)$ have definite parity, and hence it is sufficient to construct them for $\tilde{z} \geq 0$. However, for any other form of the background radial electric field--in particular for the vector form of the potential we are interested in--the solutions $\tilde h(\tilde z)$ will not have a definite parity. This implies that while searching for normalizable solutions it is necessary to check normalizability at both boundaries, $\tilde{z} = - \infty$ and $\tilde{z} = \infty$, and this complicates the search for marginally stable modes in our case. The discussed feature was not taken into account in earlier studies in Ref. \cite{Chuang:2010ku}, where the authors attempted to check the stability of the finite-baryon-density configuration of Ref. \cite{Bergman:2007wp}, hence invalidating their conclusions. The parity violation arises precisely from the presence of the Chern-Simons term. 

\subsection{Searching for marginally stable modes}
Having understood the possible source of the instabilities stemming from the presence of the Chern-Simons term, we explain here how we searched for the marginally stable modes. In the nonantipodal case the Lagrangian for marginally stable modes is still of the form \eqref{eq.LagInstability}, but now the functions $a(\tilde z)$ and $b(\tilde z)$ are more complicated than Eq. \eqref{eq.aANDbFUNCS} and also depend on $\tilde{x}_4 (\tilde{z})$. The Chern-Simons term, since it is topological, remains the same. The equations of motion following from the Lagrangian \eqref{eq.LagInstability} read 
\be
\label{eq.EOMInstability}
\tilde h''( \tilde z) + \f{a'(\tilde z)}{a (\tilde z)} \tilde h'(z) - \f{b(\tilde z) + 12 \alpha  k \tilde A'_0(\tilde z)}{a(\tilde z)} \tilde h(\tilde z) =0. 
\ee
As explained earlier, the onset of the tachyonic instability is usually signaled by a marginally stable mode that is normalizable at \emph{both} asymptotic regions. We search for such modes in two different ways. First, we fix the value of $\tilde{h}(0)$ by setting it to unity and solve Eq.\eqref{eq.EOMInstability} from $\tilde{z}=0$ to $\tilde{z} = \infty$ and from $\tilde{z}=0$ to $\tilde{z} = -\infty$ for various values of $\tilde{h}'(0)$. This approach misses solutions which have $\tilde{h}(0)=0$, so we also need to fix the value of $\tilde h'(0)$ (also to unity) and solve for the two ranges of $\tilde{z}$ for various values of $\tilde{h}(0)$. In parity-preserving Lagrangians this would correspond to looking for even and odd solutions.

\subsection{Scanning background solutions}

\begin{figure}
\includegraphics[width=8.5cm]{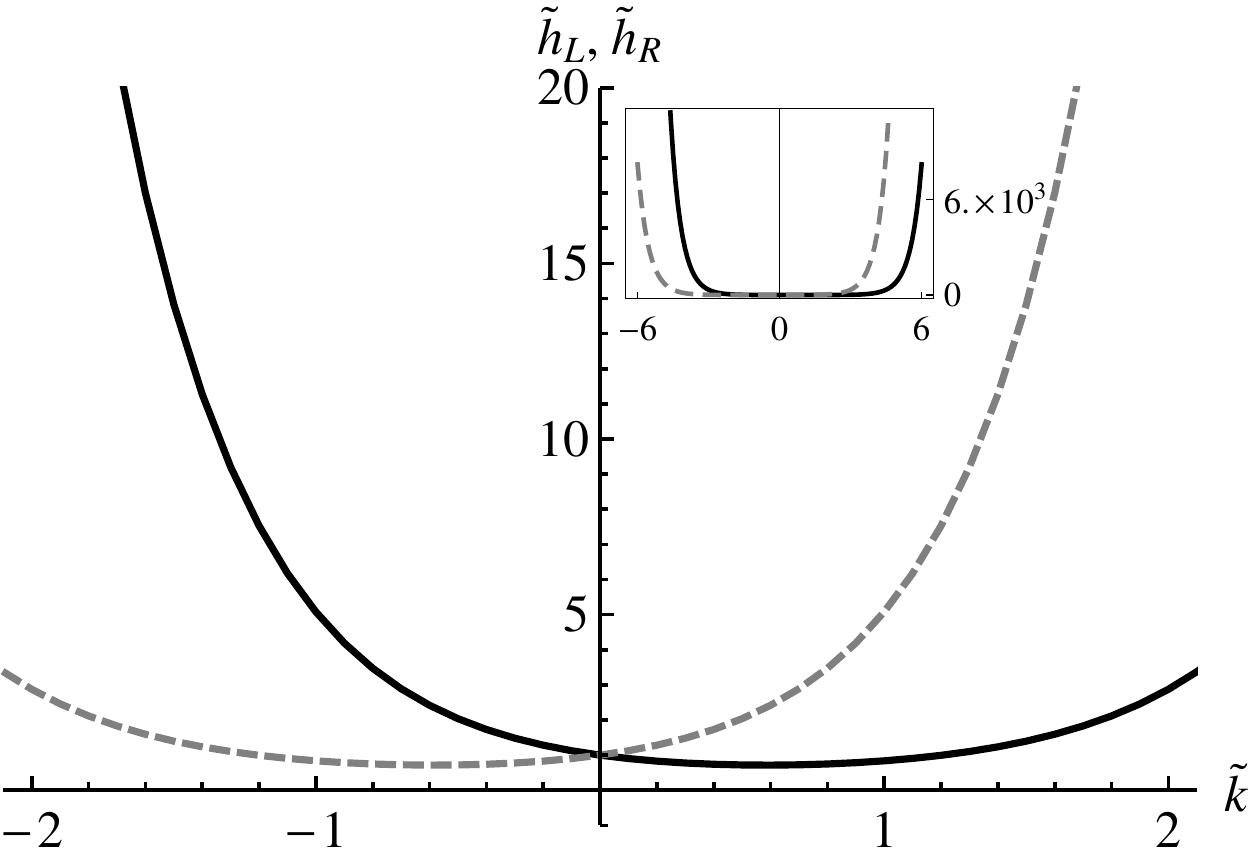}
\caption{Asymptotic values of the gauge field perturbation $\tilde{h}$ at $\tilde{z}=-\infty$ ($\tilde{h}_{L}$ given by the solid black curve) and $\tilde{z}=\infty$ ($\tilde{h}_{R}$ given by the dashed gray curve) as functions of the momentum of a mode for  baryon density $\tilde E=1000$, $\tilde{h}'(0)=0$, and antipodal embedding. No marginally stable mode is present. The inset plot shows that functions quickly achieve large values. Note that the plot is symmetric with respect $\tilde{z} \leftrightarrow - \tilde{z}$ only because $\tilde{h}'(0)=0$. }
\label{fig:MarginalModesDen1000Xi1hPrime0}
\end{figure}

\begin{figure}
\includegraphics[width=8.5cm]{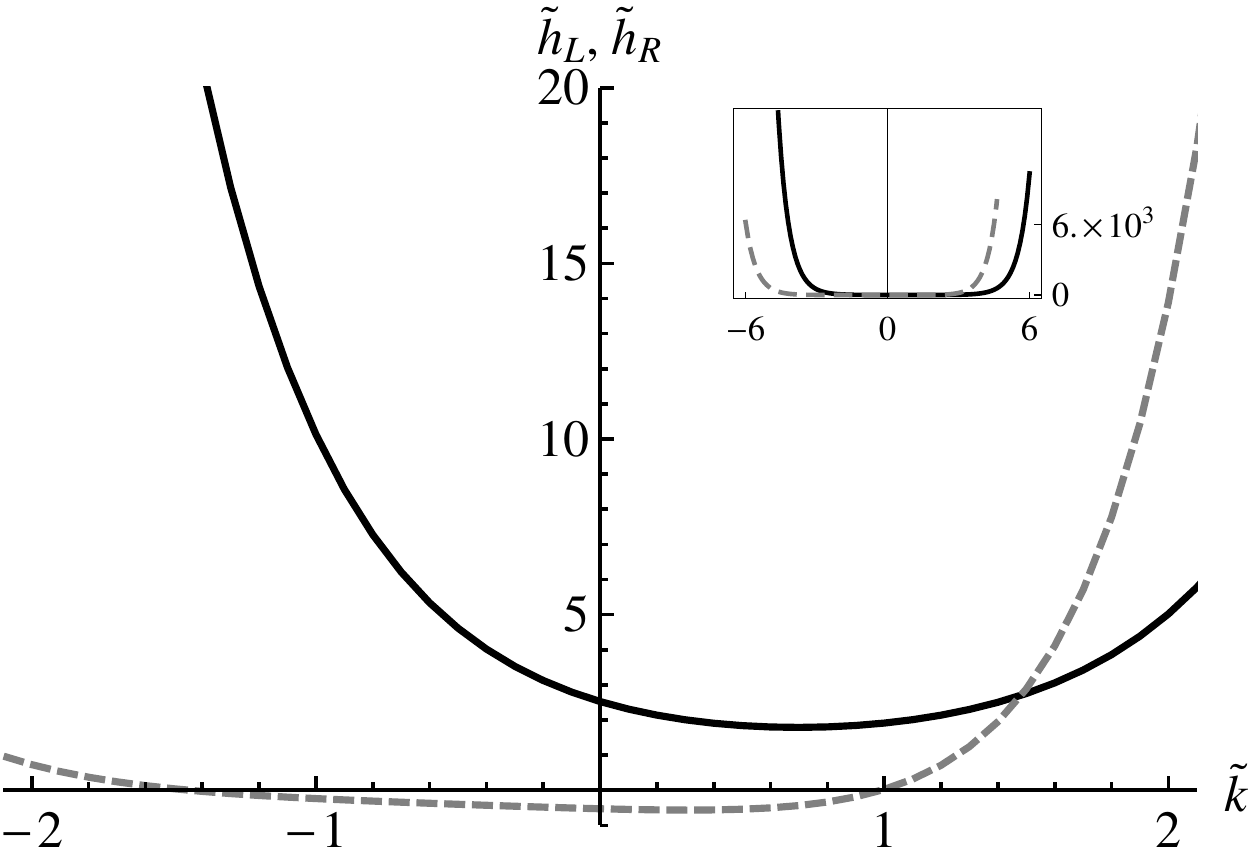}
\caption{Negative scan for the marginally stable mode at $\tilde E=1000$, $\tilde h'(0)=1$, and  antipodal embedding with the same notation as in Fig. \ref{fig:MarginalModesDen1000Xi1hPrime0}. Note that the plot is no longer symmetric with respect to $\tilde{z} \leftrightarrow - \tilde{z}$ as $\tilde h'(0) \neq 0$.}
\label{fig:MarginalModesDen1000Xi1hPrime1}
\end{figure}

\begin{figure}
\includegraphics[width=8.5cm]{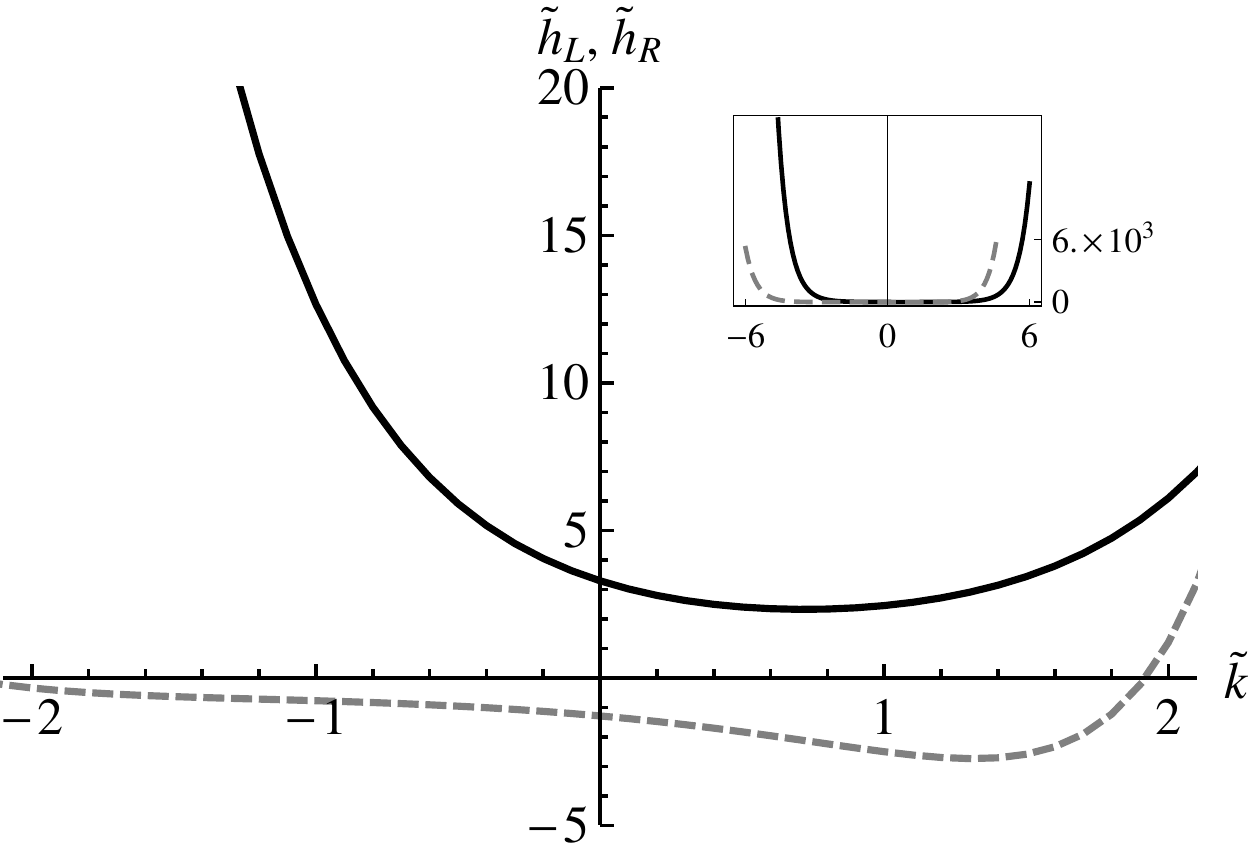}
\caption{Negative scan for the marginally stable mode at $\tilde E=1000$, $\tilde h'(0)=1.5$, and  antipodal embedding with the same notation as in Fig. \ref{fig:MarginalModesDen1000Xi1hPrime0}.}
\label{fig:MarginalModesDen1000Xi1hPrime1p5}
\end{figure}

\begin{figure}
\includegraphics[width=8.5cm]{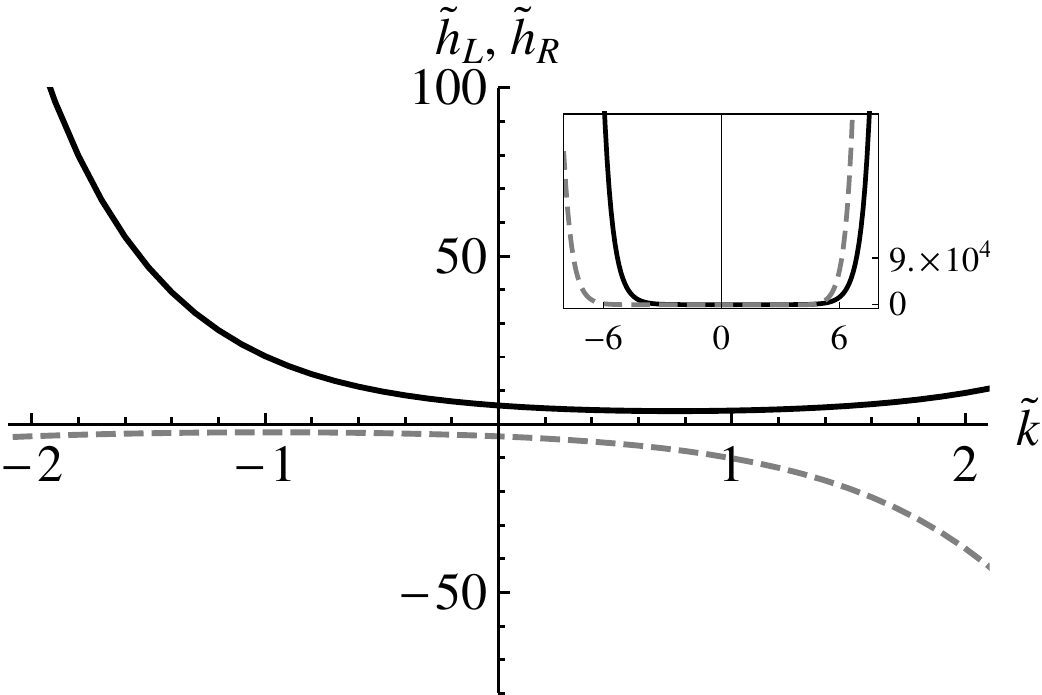}
\caption{Negative scan for the marginally stable mode at $\tilde E=1000$, $\tilde h'(0)=3$, and  antipodal embedding with the same notation as in Fig. \ref{fig:MarginalModesDen1000Xi1hPrime0}.}
\label{fig:MarginalModesDen1000Xi1hPrime3}
\end{figure}

\begin{figure}
\includegraphics[width=8.5cm]{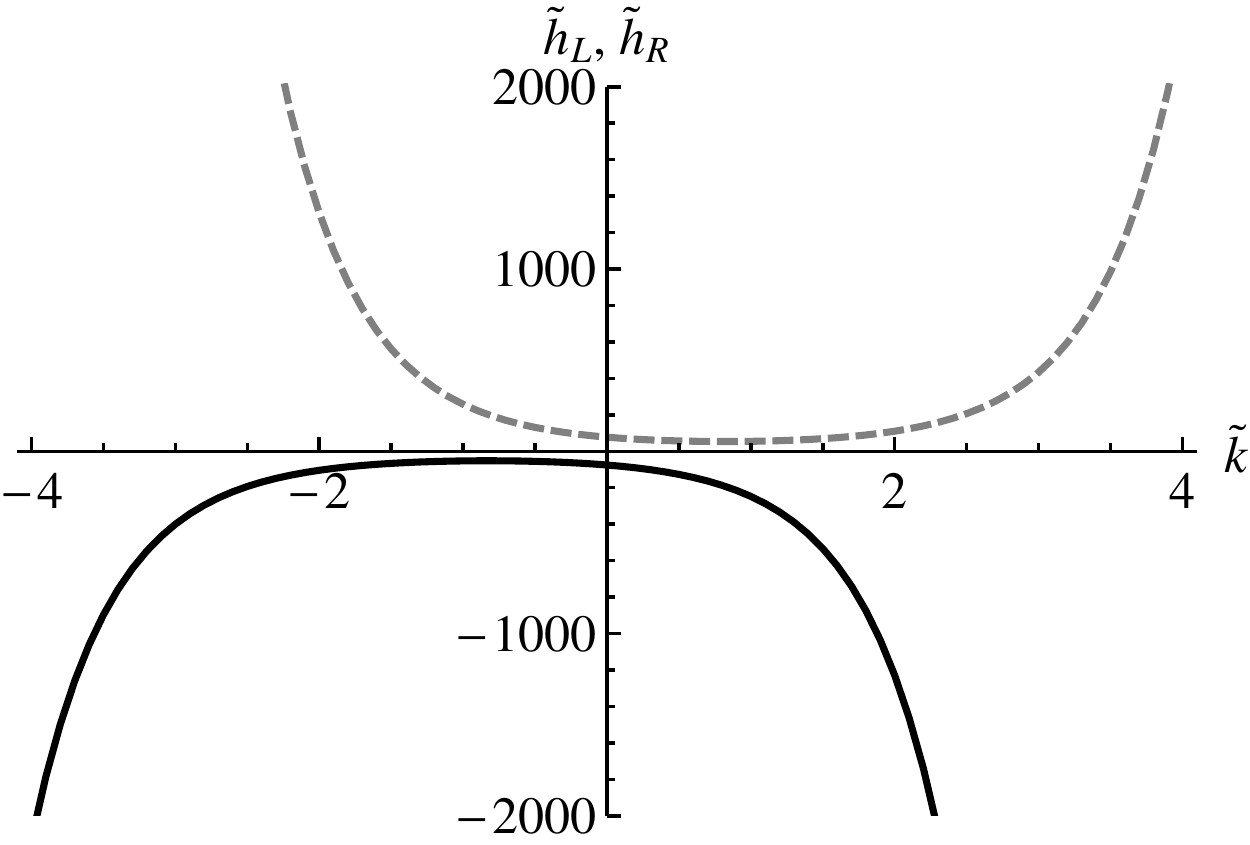}
\caption{Negative scan for the marginally stable mode at $\tilde E=1000$, $\tilde h'(0)=50$, and the antipodal embedding with the same notation as in  Fig. \ref{fig:MarginalModesDen1000Xi1hPrime0}.}
\label{fig:MarginalModesDen1000Xi1hPrime50}
\end{figure}

After thorough searches of marginally stable modes for various asymptotic separations of the D8-brane and various baryon charge densities, we conclude that in the model we study that there are no such modes because the electric field on the D8-brane is too small. 
To make this case, we plot the values of $\tilde{h}_R \equiv \tilde{h} (\infty)$ and $\tilde{h}_L \equiv \tilde{h}(-\infty)$ as functions of the momentum $\tilde k$. A normalizable mode will be one for which $\tilde{h}_L$ \emph{and} $\tilde{h}_R$ vanish simultaneously. In Figs. \ref{fig:MarginalModesDen1000Xi1hPrime0} to \ref{fig:MarginalModesDen1000Xi1hPrime50} we show the results for the antipodal case at normalization chosen so that $\tilde{h}(0)=1$ for various values of the baryon density $\tilde{E}$ and the infrared parameter $\tilde{h}'(0)$.  Note that scanning for $\tilde{h}'(0)>0$ is enough because of the symmetry of the equations of motion [flipping the sign of $\tilde{h}'(0)$, flipping the sign of $\tilde{k}$, and exchanging $\tilde{z}$ with $- \tilde{z}$]. In the inset we show the same plots but for a wider range of $\tilde{k}$.

We clearly see that the qualitative nature of the plots is the same, and with the increase of $\tilde{h}'(0)$ just the overall scale of the plots increases. As is visible in the inset plots, $\tilde{h}_L$ and $\tilde{h}_R$ become very large already for intermediate values of the momentum $\tilde{k}$. Thus, if there were marginally stable modes they would manifest themselves at relatively small values of $\tilde{k}$. Although it is possible to make the mode normalizable at one of the asymptotic regions of the D8-brane, we did not find any example where it is normalizable on both ends, neither in the antipodal nor in the nonantipodal case.

\begin{figure}
\includegraphics[width=8.5cm]{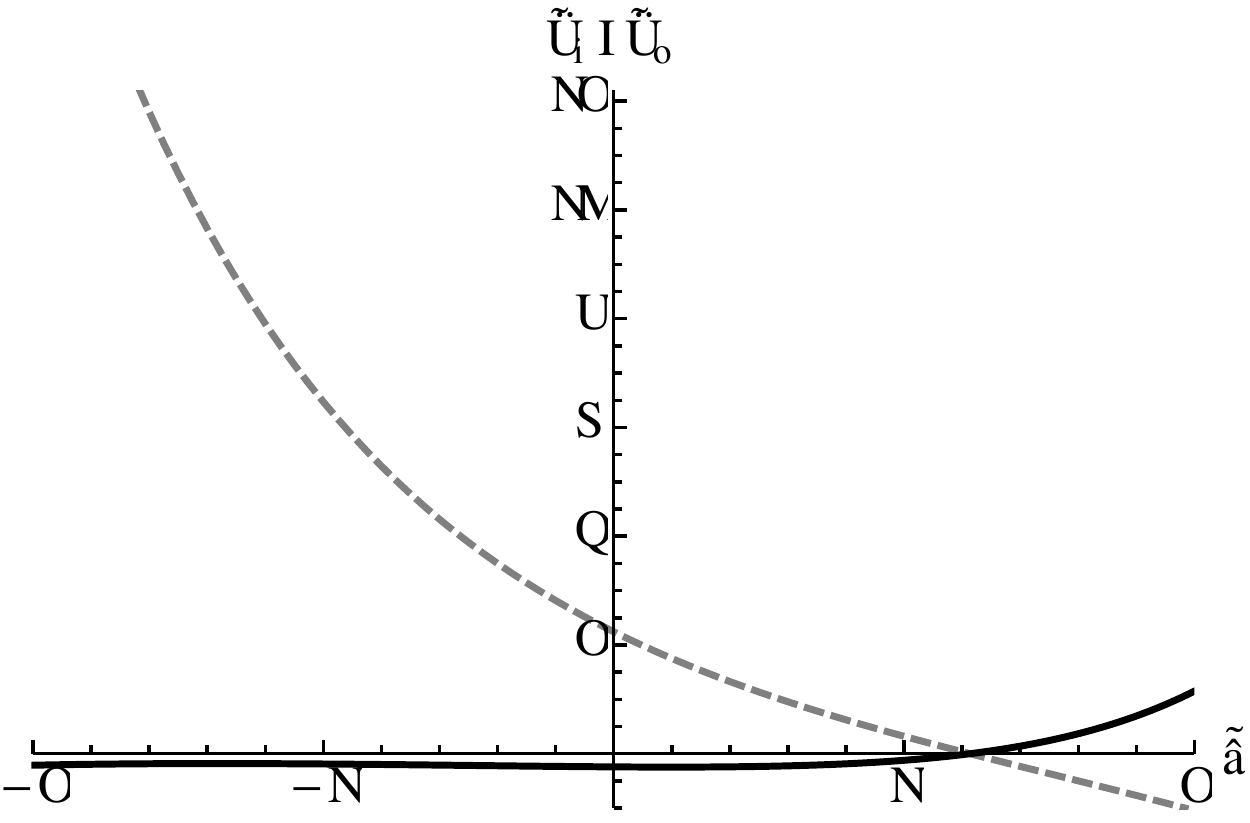}
\caption{The appearance of the marginally stable mode ($\tilde{h}_{L/R} = 0$ at $\tilde{k}\approx 1.23$) when the baryon mass is increased fivefold ($\tilde E=100$, $\tilde h'[0] \approx 1.98$, \emph{and} $\beta=20 \pi^2$). The notation is the same as in Fig.~\ref{fig:MarginalModesDen1000Xi1hPrime0}.}
\label{fig:MarginalModesDen1000Xi1hPrime3IncreasedMass}
\end{figure}

The absence of this particular marginally stable mode turns out to be a subtlety related to the smallness of the radial electric field generated by the D4-brane charge distributions on the D8-brane. Indeed, if we treat the mean-field action in the bottom-up spirit and treat parameters appearing there as free, by increasing the baryon mass ($\beta$) five-fold we do see marginally stable modes of the form \eqref{eq.InstPert}. This is clearly visible in Fig.  \ref{fig:MarginalModesDen1000Xi1hPrime3IncreasedMass}. As our mean-field description makes various assumptions, it is entirely conceivable that microscopic parameters, such as the baryon mass or the assumed DBI form of the kinetic term for the gauge field, do not enter the mean-field action directly but rather get renormalized in the process of coarse-graining. We comment on this possibility in the Conclusions. Note also that the absence of this particular marginally stable mode is not enough to claim that the system in question is linearly stable.

\section{Vector and axial meson spectra at large densities \label{sec.mesons}}

\subsection{Mesons at nonzero baryon density}
The key conceptual issue that we are interested in is understanding whether the chiral symmetry can be restored at low temperatures by finite-density effects. In the single-flavor case that is being considered in this paper, the analogue of the chiral symmetry is the axial symmetry. One of the ways  the chiral symmetry breaking would manifest itself in our case is if the parity-odd and the parity-even mesons fail to form doublets. Indeed, in the spin-one sector that we focus on in this article, the axial and vector meson masses differ. For the lowest two mesons the relative mass difference \eqref{eq.Deltam} is as large as $40\%$, which can be seen directly in Fig. \ref{fig.Deltam}.

At nonzero density there are two effects which might influence the spectrum of mesons, and both require parametrically large densities of order ${\cal O}(\lambda^{2})$. For the first one, the DBI action is nonlinear and the kinetic term for the gauge field perturbations gets modified by the presence of the electric field on the D8-brane provided the latter is big enough, i.e., ${\cal O}(\lambda)$. This exactly corresponds to baryon densities of order ${\cal O}(\lambda^{2})$. We are not aware of any work that addresses this issue for the electric field sourced by the four-dimensional holographic baryon densities, which is the relevant dual description of order ${\cal O}(\lambda^{2})$ baryon densities. However, as a similar effect is also present in the case of a large enough density of the axial charge, 
it does not seem to be of particular relevance when looking for possible holographic realizations of the quarkyonic phase.

The second effect, which to our knowledge has not been explored in the literature, are interactions of gauge field perturbations corresponding to the axial and the vector mesons with the holographic density of the baryons. Such a possibility follows from the parametrically large baryon density that we consider.

In the following, we will analyze the influence of both of these effects on the masses of the lowest (transverse) vector and axial mesons.

\subsection{The relevance of direct interactions of the holographic mesons and baryons}
Let us start with the analysis of how various terms in the action scale with $N_{c}$ and $\lambda$. This will determine the parametric regime in the baryon density, in which the meson-baryon interactions significantly influence the vector and axial meson spectrum. Note that in this subsection we will be operating on the original, i.e., \emph{nonrescaled} quantities, as we will be discussing the microscopic picture.

The mesons of interest are perturbations of the U(1) gauge field on the world volume of the D8-brane. The kinetic term for the gauge field, after expanding the DBI action to quadratic order, has a factor of $N_{c} \lambda$ in front\footnote{This can be seen from Eq. \eqref{eq.SDBI} after Kaluza-Klein reduction on $S^{4}$ and expanding it to quadratic order in the field strength.}. In the vacuum there are no other terms that can compete with it at quadratic order in the gauge field perturbations $\delta A$, and hence the overall scaling is irrelevant for the masses of the mesons. At nonvanishing baryon density, the holographic baryons will couple to the gauge field perturbations. Then the factor of (the inverse of) $N_{c} \lambda$ will control the strength of the electromagnetic interactions with \emph{single} baryons; schematically,
\be
\label{eq.scalingofMaxwelleqns}
\partial^2 \delta A = \frac{1}{N_{c} \lambda} \delta j,
\ee
where $\delta j$ is the current made out of the holographic baryons. This equation is nothing but an approximate and simplified version
of Eq. \eqref{eq.DBIgeneralizationofMaxwell}.
As the baryon charge scales as $N_c$, the current $\delta j$ schematically reads
\be
\label{eq.scalingofcurrent}
\delta j = N_{c} n \delta u.
\ee
We assume that the perturbations of interest do not affect the number density $n$, which will indeed turn out to be the case.

As meson perturbations are, in the single-flavor case, the electromagnetic waves on the D8-brane, there will be a Lorentz force acting on the holographic baryons.  We can schematically write Newton's second law $ma=qE$ as
\be
\label{eq.scalingof2ndlaw}
N_{c} \lambda \partial \delta u = N_{c} \partial \delta A,
\ee
where $N_{c} \lambda$ on the lhs comes from the baryon mass, whereas $N_{c}$ and $\partial \delta A$ come respectively from the baryon charge and the electric component of the electromagnetic wave on the D8-brane. Although, as follows from Eq. \eqref{eq.scalingof2ndlaw}, baryon acceleration\footnote{If there is an electric field associated with the perturbation.} and hence baryon velocity\footnote{Assuming oscillations with O(1) frequency, as is the case with meson masses in vacuum.} will be $1/\lambda$-supressed,
\be
\delta u = O(1/\lambda),
\ee
and for ${\cal O}{(\lambda^{2})}$ number densities $n$ the current $\delta j$ will contribute to the equation for meson perturbations\eqref{eq.scalingofMaxwelleqns} at the same order of large-$N_{c}$ and large-$\lambda$ scaling as the kinetic term. This implies that the electromagnetic response of baryons may significantly affect the propagation of electromagnetic waves on the D8-brane and hence the meson spectrum\footnote{One can easily see that the form of the equations \eqref{eq.scalingofMaxwelleqns}--\eqref{eq.scalingof2ndlaw} is not affected by simultaneous rescalings of $\delta A$ and $\delta u$ by a factor of $\lambda$. This is the reason why meson-baryon interactions of the form \eqref{eq.scalingofMaxwelleqns}--\eqref{eq.scalingof2ndlaw} are captured in a precise fashion by the rescaled action \eqref{eq.Seff} + \eqref{eq.Sdustresc}. Although there we were arguing that we shall never be interested in studying macroscopic velocities in spacelike directions, because of the scaling argument elucidated above, the action \eqref{eq.Seff} + \eqref{eq.Sdustresc} can nevertheless be used to (partly) calculate the equations of motion for meson perturbations in the presence of interactions with baryons.}.

One can clearly see that in the case of the marginally stable modes studied in the previous section, there will be no Lorentz force acting on the baryons. This is because baryons are at rest in equilibrium and the marginally stable mode \eqref{eq.InstPert} does not generate an electric field. This justifies neglecting the dynamics of charges in the analysis of the previous section. When it comes to the meson perturbations, these necessarily introduce an electric field in one of the spacelike directions. Such an electric field will accelerate charges according to Eq.\eqref{eq.scalingof2ndlaw}. However, once one of the charges is displaced, one may wonder whether there is no restoring force generated from interactions with other charges from the lattice. We can estimate the magnitude of the restoring force by expanding the interaction energy of a single baryon with others [with the notation the same as in Eq. \eqref{eq.deltaeps}] up to quadratic order in its displacement $\delta x$,
\bea
&&(\text{restoring force}) \cdot \delta x = \nonumber\\
&&\sum_{\delta V} \frac{N_{c}}{\lambda} \frac{1}{d^{2} \left( p_{1}^{2} + p_{2}^{2} + (p_{3} + \frac{\delta x}{d})^{2} + p_{4}^{2} \right)},
\eea
leading to
\be
\label{eq.restoringforce}
\text{restoring force} \sim - \frac{N_{c}}{d^{4} \lambda } \delta x \approx - N_{c} \lambda \, \delta x.
\ee
This implies that Eq. \eqref{eq.scalingof2ndlaw} needs to be supplemented with the additional term \eqref{eq.restoringforce}. Taking into account that $\delta u = \dot{\delta x}$, we end up with the schematic expression \emph{resembling} a forced harmonic oscillator,
\be
\label{eq.forcedharmonicoscillator}
N_{c} \lambda (\ddot{\delta x} + \omega_{0}^{2} \delta x) = N_{c} \dot{\delta A}.
\ee
Note that for velocities $O(1/\lambda)$ (and hence for displacements of the same order), both sides of Eq. \eqref{eq.forcedharmonicoscillator} are of the same order in $\lambda$, which means that the lattice structure does not prevent the holographic baryons from being influenced by electromagnetic waves representing the holographic mesons. As all factors of $\lambda$ cancel in both Eq. \eqref{eq.forcedharmonicoscillator} and Eq. \eqref{eq.scalingofMaxwelleqns}, we conclude that interactions of the baryonic medium with the mesons will influence the properties of the latter.

At this moment we do not have a precise form of Eq. \eqref{eq.forcedharmonicoscillator}, not to mention the precise value of the coefficient $\omega_{0}$. In particular, note that if $\delta x$--denoting the displacement in one of the spacelike field theory directions--depends only on time then there is no restoring force, as the lattice as a whole moves uniformly. This implies that $\delta x$ must enter the restoring-force formula \eqref{eq.restoringforce} with the derivative in the radial direction, which is the reason why Eq. \eqref{eq.forcedharmonicoscillator} only resembles a forced harmonic oscillator. Note also that such a term might appear in the effective description \eqref{eq.Sdustresc} as a derivative correction.

In the following, we will explore the implications of the interactions of the mesons with the baryons while completely neglecting the restoring force. This could be justified if it turned out that $\omega_{0}$ in the cases of interest was much lower than the mass of the lowest vector or axial meson. We do not know at the moment whether or not this is true. Our motivation is rather to show that density-enhanced holographic meson-baryon interactions may lead to very interesting physical effects, which provides a  very strong motivation for more detailed future explorations of the phase diagram of the Sakai-Sugimoto model at ${\cal O}{(\lambda^{2})}$ baryon density.

\subsection{Gauge field perturbations describing vector and axial mesons at ${\cal O}{(\lambda^{2})}$ baryon density}
In the following, we will search for the normalizable gauge field perturbations on the D8-brane of the form \eqref{eq.mesonperturbation}. As explained in Sec. \ref{subsec.vacuumsol}, the ansatz \eqref{eq.mesonperturbation} corresponds to zero-momentum axial and vector meson currents in the $\tilde{x}^{3}$ direction. Their frequencies have a direct interpretation in terms of meson masses. In the vacuum, the differences in the masses of the corresponding vector and axial mesons come from the broken axial symmetry\footnote{A very interesting question that we are not going to analyze in this article is the fate of the holographic mode corresponding to the pion ($\eta'$ meson) at parametrically large baryon densities. If the pion remains massless, at zero momentum its frequency should vanish. The pion corresponds to a perturbation of the form $A_{\mu} = \psi(\tilde{z}) \partial_{\mu} \phi(x)$, where $\psi(\tilde{z})$ is antisymmetric and approaches nonzero and opposite constant values on the boundaries \cite{Sakai:2004cn}. At zero momentum, i.e. $\phi(x) = e^{- \imath \tilde{\omega} \tilde{x}^{0}}$, such a mode generates an oscillating radial electric field and, as our system has broken translational invariance in the radial direction, studies of this mode in the presence of the dense baryonic medium seem to be more difficult than that of the modes corresponding to vector and axial mesons at zero momentum. If there is a nonzero frequency for which this mode exists, then the pion effectively becomes massive, which would suggest that the chiral (axial) symmetry might be restored. It would be very interesting to analyze this issue in more detail, but the following argument suggests that the pion will remain massless: if the spectrum does not jump discontinuously as we increase the density, the pion will still correspond to a perturbation of the form $A_{\mu} = \psi(\tilde{z}) \partial_{\mu} \phi(x)$. This perturbation does not change under $\phi(x) \rightarrow \phi(x)+{\rm const}$, and therefore this must be a symmetry of the system, which in turn can only be the case if the pion is massless.}.

Here we will reexamine the perturbations \eqref{eq.mesonperturbation}, but now in the presence of a parametrically large baryon density, as was considered in Sec. \ref{subsec.vacuumsol}. The goal is to obtain the frequencies of the normalizable modes, which we understand as masses of mesons in the dense baryonic medium, and compare them with the masses of the corresponding mesons in the vacuum. If the chiral (axial) symmetry is (approximately) restored, then the masses of the corresponding axial and vector mesons will be (approximately) equal.

The notion of chiral symmetry restoration, which our description possibly allows for, is the one in which perturbations on the $\overline{D8}$-brane and D8-brane (approximately) decouple because of the interactions with the holographic baryons.

Indeed, for perturbations of the form \eqref{eq.mesonperturbation} this might be in principle possible, as the ansatz \eqref{eq.mesonperturbation} leads to a time-dependent electric field in $\tilde{x}^{3}$ direction. The electric field accelerates baryons leading to a time-dependent, but uniform, current in the  $\tilde{x}^{3}$ direction. This current backreacts on the electric field through the standard electromagnetic coupling present in the action \eqref{eq.Sdustresc}, which (intuitively) lowers the original electric field, in complete analogy with the London equation in the case of an electromagnetic field trying to penetrate the interior of a superconductor. The full picture will turn out to be more intricate because of the $\tilde{z}$-dependent background geometry and the charge density.

Of course, in contrast to the black hole embedding when the chiral symmetry is exactly restored, for mesons of sufficiently large mass the holographic baryon density will not be very relevant. We will revisit this issue in the following subsection by doing a Schr\"{o}dinger-equation analysis of meson perturbations searching for the potential barrier emerging from bulk meson-baryon interactions and in this subsection we will present numerical results of the lowest part of the vector and axial meson spectrum as a function of the asymptotic separation of the D8- and $\overline{D8}$-branes and the baryon density $B$.

The equations of motion for the perturbation \eqref{eq.mesonperturbation} will follow from taking the functional derivative of the action \eqref{eq.Seff} + \eqref{eq.Sdustresc} with respect to $\delta\tilde{a}_{3}(\tilde{t},\tilde{z})$ and $\delta \tilde{u}_{3}(\tilde{t},\tilde{z})$. After going to frequency space, the equation for $\delta\tilde{a}_{3}(\tilde{z})$ takes the schematic form
\bea
&&\delta \tilde{a}_{3}''(\tilde{z}) + C_{1}(\tilde{z}) \delta \tilde{a}_{3}'(\tilde{z}) + \tilde{\omega}^{2} C_{0}^{(\omega)}(\tilde{z}) \delta \tilde{a}_{3}(\tilde{z}) +
\nonumber\\
&&\tilde{w}^{2} C_{0}^{(w)}(\tilde{z}) \delta \tilde{u}_{3}(\tilde{z}) = 0,
\eea
where the exact form of $C_{0}^{(w)}(\tilde{z})$, $C_{0}^{(\omega)}(\tilde{z})$ and $C_{1}(\tilde{z})$ is not very illuminating and can be found in the Appendix\footnote{Actually, we skip the form of $C_{0}^{(w)}(\tilde{z})$, as in the master equation \eqref{eq.mesonperturbationsinmedium} it appears in a combination with other terms and in the Appendix we only provide the form of the resulting expression, denoted by $C_{0}(\tilde{z})$.}. Note also the presence of the term $\tilde{w}^{2} C_{0}^{(w)}(\tilde{z})$, which was absent in the vacuum case and here encodes the interactions of mesons with baryons. This term will lead to new physical effects.

The equation for $\delta\tilde{u}_{3}(\tilde{z})$ takes the form of Eq. \eqref{eq.eomsfordust4ua},
\be
\gamma \tilde{w}(\tilde{z})^2 \, \delta \tilde{a}_{3}(\tilde{z}) = - 2 \tilde{\lambda}(\tilde{z})\, \delta \tilde{u}_{3}(\tilde{z})
\ee
with $\tilde{\lambda}$ given by
\be
\tilde{\lambda} = \frac{1}{2} \, \beta \, \tilde{\xi}^{1/4}(1 + \frac{\tilde{z}^{2}}{\tilde{\xi}^{2}})^{1/12} \tilde{w}^{2}.
\ee
The formula above is the curved-space analogue of the London equation and leads to a decoupled equation for the meson perturbation,
\bea
\label{eq.mesonperturbationsinmedium}
&&\delta \tilde{a}_{3}''(\tilde{z}) + C_{1}(\tilde{z}) \delta \tilde{a}_{3}'(\tilde{z}) + \omega^{2} C_{0}^{(\omega)}(\tilde{z}) \delta \tilde{a}_{3}(\tilde{z}) +
\nonumber\\
&& C_{0}(\tilde{z}) \delta \tilde{a}_{3}(\tilde{z}) = 0.
\eea

The form of the coefficient $C_{0}(\tilde{z})$ can be found in the Appendix and together with $C_{1}(\tilde{z})$ and $C_{0}^{(\omega)}(\tilde{z})$ it encodes the form of the embedding, the background electric field and the interactions of mesons with baryons. It can be easily verified that the equation for meson perturbations is symmetric with respect to $\tilde{z} \leftrightarrow - \tilde{z}$ and hence it is sufficient to solve it for $\tilde{z}$ between $0$ and $\infty$. Imposing Dirichlet boundary conditions at infinity leads to the spectrum of mesons at ${\cal O}{(\lambda^{2})}$ densities. 

\subsection{Meson spectra and the effective potential for meson perturbations\label{subsec.SchrodingerPotential}}

The presence of density-enhanced direct holographic meson-baryon interactions changes the form of the equation for meson perturbations, and hence influences the spectrum of the axial and vector mesons. In order to better understand the features introduced by the interactions in question, it will be extremely useful to rewrite the equation for perturbations \eqref{eq.mesonperturbationsinmedium} in the Schr\"{o}dinger form, i.e.,
\be
\label{eq.Schroedinger}
-\frac{1}{2} \frac{d^{2}}{d\sigma^{2}} \Psi(\sigma) + V(\sigma) \Psi(x) = E \Psi(\sigma),
\ee
where the energy $E$ of the wave function $\Psi$ is the square of the rescaled frequency $\tilde{\omega}$ and hence is proportional to the square of the meson mass $m$ \eqref{eq.massintermsofomega},
\be
E = \tilde{\omega}^2 = \frac{9}{4} m^{2}/M_{KK}^{2},
\ee
$\Psi(\sigma)$ is related to $\tilde{a}_{3}(\tilde{z})$, and $\sigma$ is a new radial coordinate that leads to the canonical kinetic term in Eq.\eqref{eq.Schroedinger}. Simple manipulations of \eqref{eq.mesonperturbationsinmedium} allow us to obtain the potential term $V$. For simplicity, $V$ can be expressed as a function of $\tilde{z}$ instead of $\sigma$,
\bea
V_{Schr}(\tilde{z}) =&& - \frac{C_{0}(\tilde{z})}{C_{0}^{(\omega)}(\tilde{z})} + 
\frac{C_{1}(\tilde{z})^{2}}{4 C_{0}^{(\omega)}(\tilde{z})} - 
\frac{{5 C_{0}^{(\omega)}} ' (\tilde{z})^{2}}{16 C_{0}^{(\omega)} (\tilde{z})^{3}} + \nonumber\\
&&\frac{C_{1}'(z)}{2 C_{0}^{(\omega)}(\tilde{z})} +
\frac{{C_{0}^{(\omega)}}''(\tilde{z})}{4 C_{0}^{(\omega)}(\tilde{z})^{2}},
\eea
where $C_{0}(\tilde{z})$, $C_{0}^{(\omega)}(\tilde{z})$, and $C_{1}(\tilde{z})$ are as in Eq. \eqref{eq.mesonperturbationsinmedium}. 

Because of the way the background spacetime is warped, $V$ needs to have the form of an infinite potential well. This feature gives rise to an infinite tower of bound states, which are simply the vector and axial mesons. However, the presence of a nontrivial embedding, nonzero background electric field, and most importantly holographic interactions of the baryons with the mesons may lead to new structure of the potential having a direct impact on the spectrum of the axial and vector mesons.

In the following we will analyze in detail two cases, which--based on more exhaustive studies--we think are representative ones. 

\begin{figure}
\includegraphics[width=8.5cm]{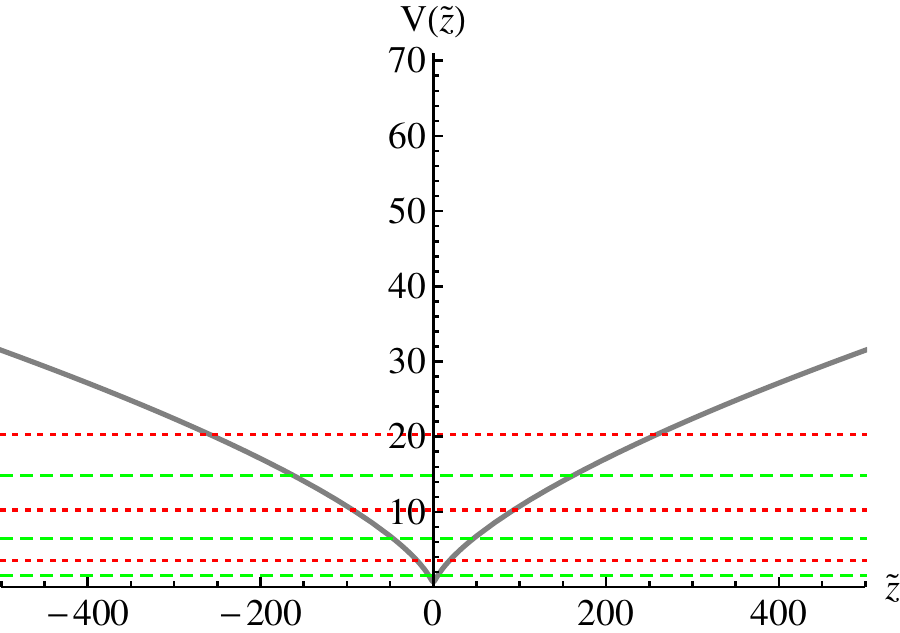}
\includegraphics[width=8.5cm]{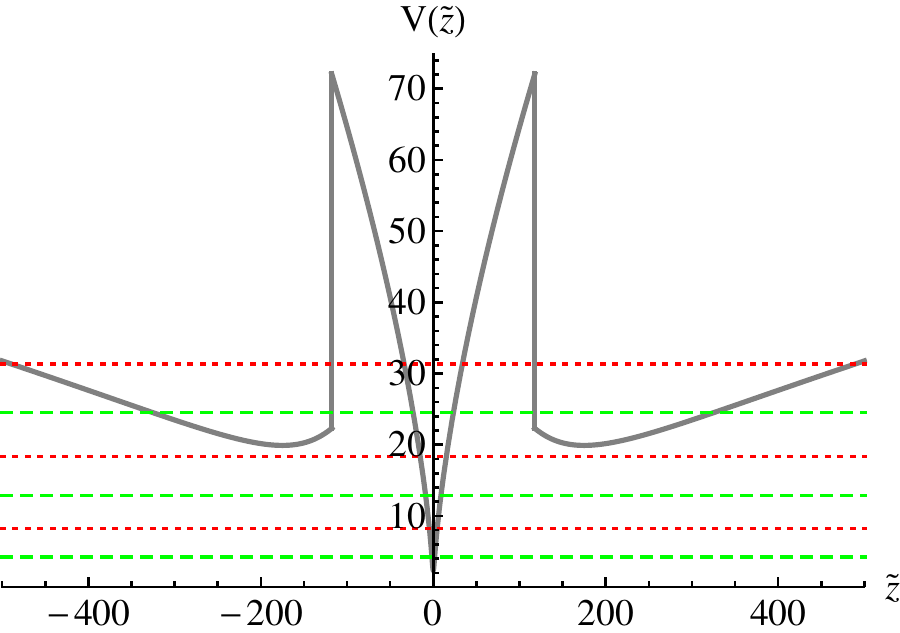}
\caption{The Schr\"{o}dinger potential and the three lowest axial and vector mesons in the antipodal case ($L/\delta\tau = 0.5$) at vanishing (top plot) and large (bottom plot, $\tilde{E} = 1000$) density. Green dashed and red dotted lines correspond to the squares of masses of the lowest three vector and axial mesons, respectively. The most interesting finding is the appearance of an emergent potential well due to the holographic baryons  on the D8-brane. It can be clearly seen that in the presence of a dense baryonic medium, the lowest axial and vector mesons are bound states not of the master potential well--which is present also in the vacuum--but rather of the emergent one. This leads to a tempting connection with a quasiparticle picture in which some of the lowest in-medium mesons are collective excitations of the  underlying Fermi surface.}
\label{fig.spectrum_antipodal}
\end{figure}

Figure \ref{fig.spectrum_antipodal} shows the Schr\"{o}dinger potential in vacuum (upper plot) and at very large density ($\tilde{E} = 1000$, lower plot) for the antipodal embedding. The green dashed and red dotted lines denote the squares of vector and axial meson masses for the three lowest mesons. As anticipated earlier, the potential has the structure of a well supporting infinitely many bound states. In the vacuum case, the potential is simple and does not have any  finer structure than the master infinite potential well coming from the warping of the target spacetime. 

If one looks at the situation at nonzero density, one sees that the potential develops substructure in the form of another, finite potential well, which is entirely due to the presence of charges on the flavor brane. The immediate observation is that the lowest vector and axial vector mesons then arise as bound states of the \emph{emergent} potential well, rather than the original one. This strongly suggests the analogy with the quasiparticle picture in which low-energy long-lived excitations, when expressed in terms of the underlying microscopic description, turn out to be collective modes of a system. Note also that the excitations are automatically color singlets, as is the case in the quarkyonic phase.

\begin{figure}
\includegraphics[width=8.5cm]{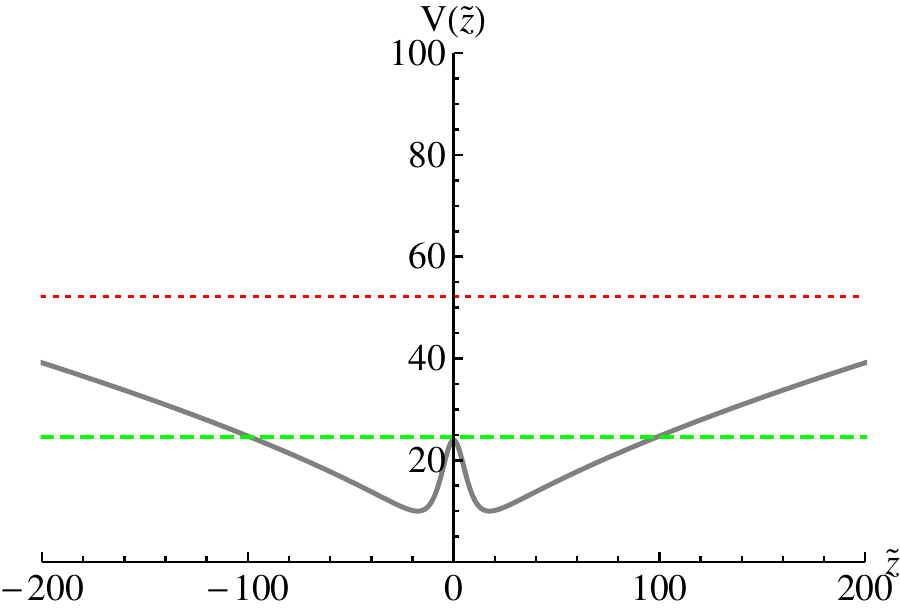}
\includegraphics[width=8.5cm]{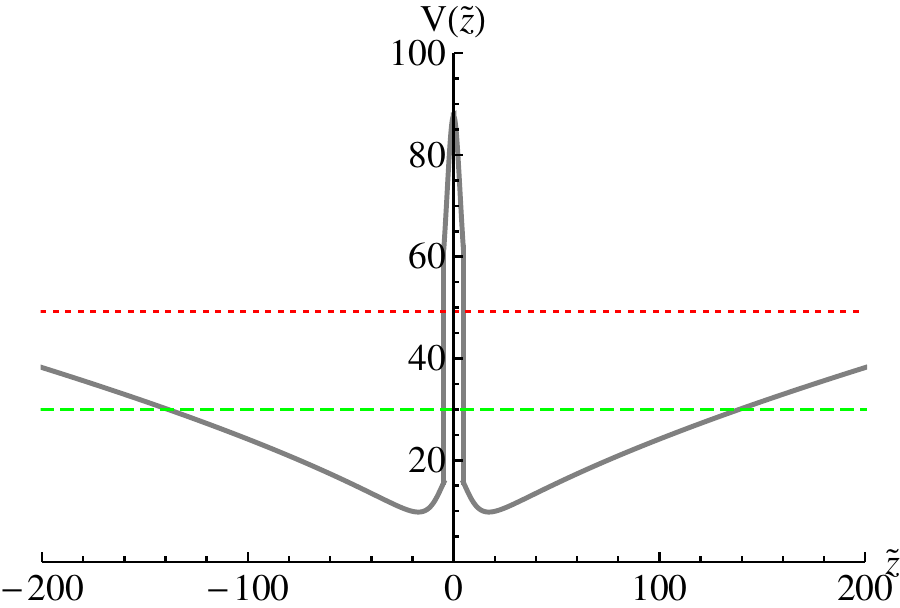}
\includegraphics[width=8.5cm]{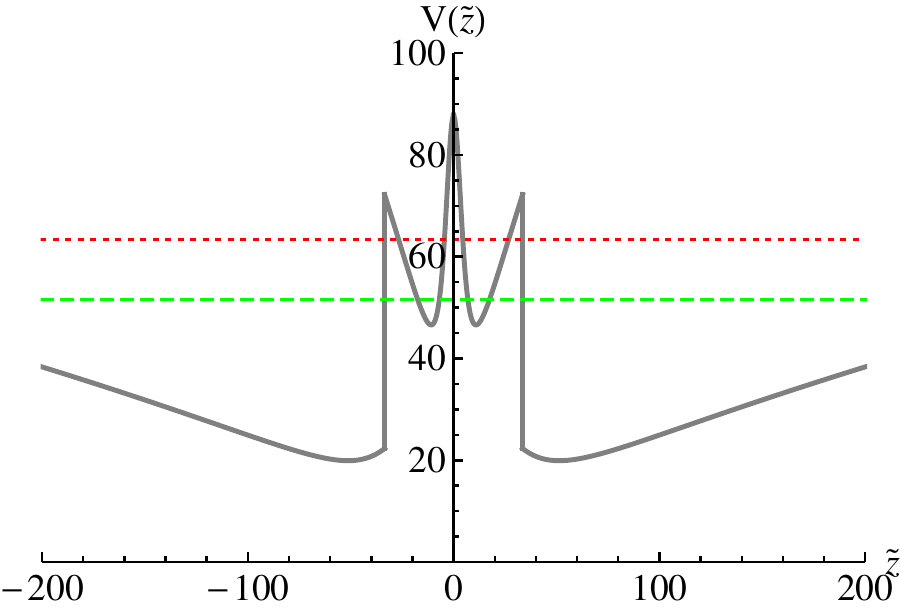}
\caption{The Schr\"{o}dinger potential and the three lowest axial and vector mesons for a very nonantipodal embedding ($L/\delta\tau = 0.05$) at zero (top plot), medium (middle plot, $\tilde{E} = 100$), and high (bottom plot, $\tilde{E} = 1000$) density. Because of the small potential barrier residing in the vicinity of $\tilde{z} = 0$, the emergent potential well does not give rise to bound states and effectively acts as a potential barrier. On the other hand, the emergent barrier leads to an approximate chiral symmetry restoration for the two lowest states, as their relative mass difference [given by Eq. \eqref{eq.Deltam}] reduces from about $40\%$ in the vacuum to $10\%$ in this case.}
\label{fig.spectrum_nonantipodal}
\end{figure}

\begin{figure}
\includegraphics[width=8.5cm]{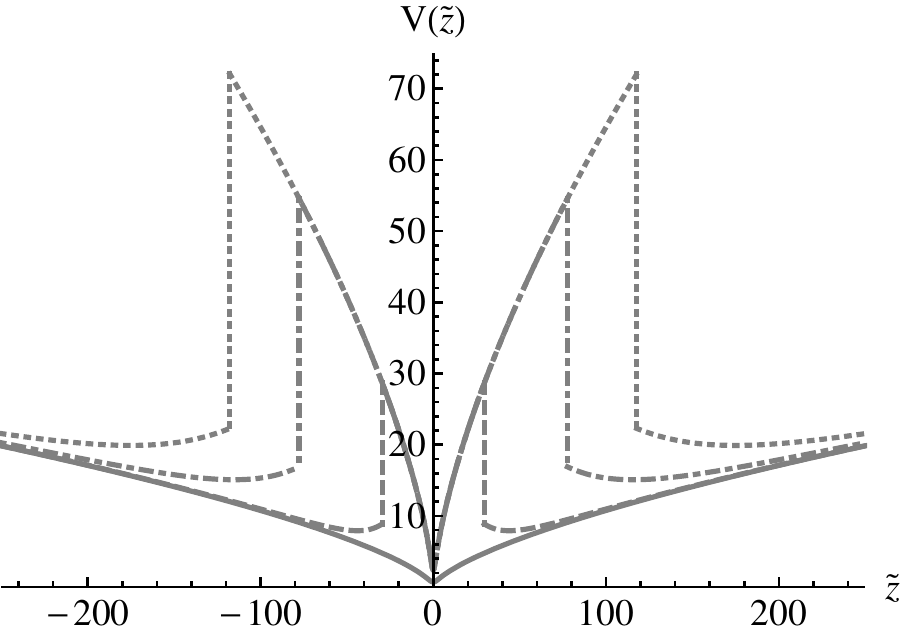}
\caption{The Schr\"{o}dinger potential in the antipodal case in the vacuum (solid curve) and at densities $\tilde{E} = 100$ (dashed curve), $\tilde{E} = 500$ (dotdashed curve), and $\tilde{E} = 1000$ (dotted curve). One clearly sees that the emergent potential barrier arising because of the charges on the D8-brane affects more and more of the lowest lying axial and vector mesons as the density increases.}
\label{fig.Vschr_as_a_func_od_density}
\end{figure}

\begin{figure}
\includegraphics[width=8.5cm]{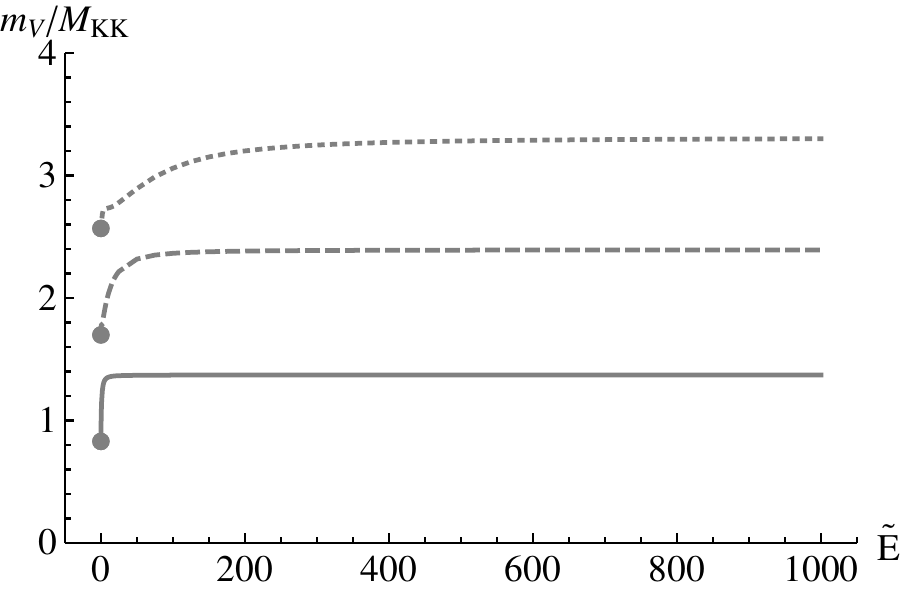}
\includegraphics[width=8.5cm]{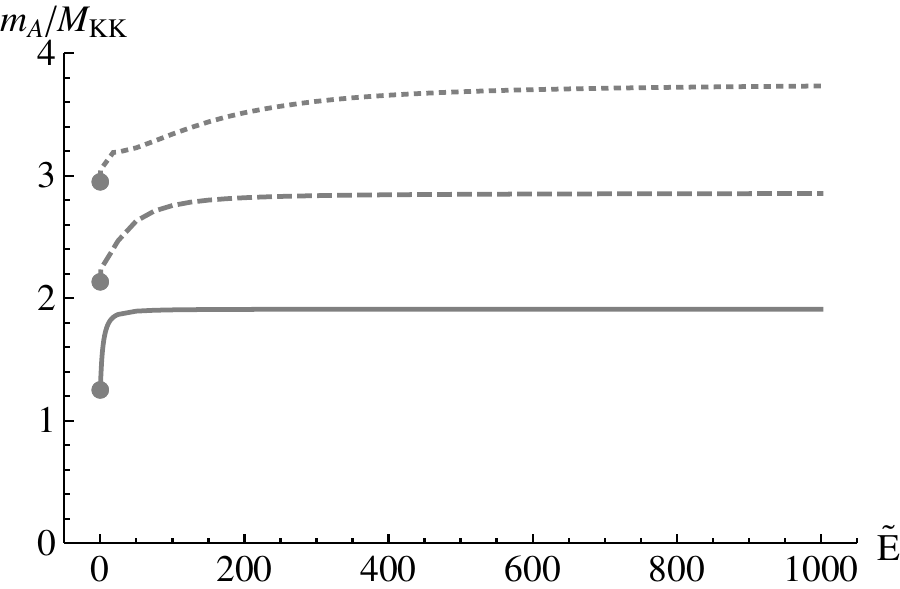}
\caption{Masses of the three lowest vector (top) and axial mesons (bottom) as a function of the rescaled baryon density $\tilde{E}$ for the antipodal embedding of the D8-brane. One can see that within the adopted approximations the masses obtain significant contributions from interactions with the dense baryonic medium. The saturation of masses with density follows from the fact that the charge distribution is not altered by adding new layers, which leads to the Schr\"{o}dinger potential, as in Fig. \ref{fig.Vschr_as_a_func_od_density}.}
\label{fig.spectrum_as_a_func_of_E}
\end{figure}


This becomes apparent in Fig. \ref{fig.Vschr_as_a_func_od_density}, which shows the Schr\"{o}dinger potential in the antipodal case for the vacuum and three representative charge densities, and in Fig. \ref{fig.spectrum_as_a_func_of_E}, which shows the masses of the the three lowest axial and vector mesons as functions of density, also in the antipodal embedding.

We are not aware of any similar results in the gauge-gravity duality, but we also want to stress at this point that our studies are preliminary and clearly--as explained in the beginning of this section--do not take into account all in-medium phenomena.

Setting this aside, we also want to point out another possibility that the density-enhanced meson-baryon interactions offer, namely that the emergent potential well from the point of view of the master potential well might also act as a potential barrier. This clearly does not happen in the antipodal case, where the emergent potential well supports a few bound states, but the asymptotic separation of the $\overline{D8}$- and D8-brane $L$ is a free parameter and we can investigate whether the situation changes when we set it to some different value.

It turns out, that in the nonantipodal case in the vacuum there is already a small potential barrier in the middle of the master potential well, as seen in Fig. \ref{fig.spectrum_nonantipodal} (top). Again, the charge on the D8-brane will result in the creation of an emergent potential well, but now this small potential barrier will not allow for any bound states inside it.
It is thus a genuine potential barrier, which offers 
the possibility of forming chiral doublets--(almost) degenerate states residing on the two sides of the barrier. Indeed, looking at the two lowest states (higher excitations have energies larger than the height of the barrier) we see that the relative ratio of the masses of axial and vector mesons \eqref{eq.Deltam}--which, as shown in Fig. \ref{fig.Deltam}, in the vacuum is of order of $40\%$--in this case can become as small as $10\%$. We interpret this result as approximate chiral symmetry restoration in the lowest part of the spectrum. Clearly, for higher excitations the potential barrier plays a less important role.

\begin{figure}
\includegraphics[width=8.5cm]{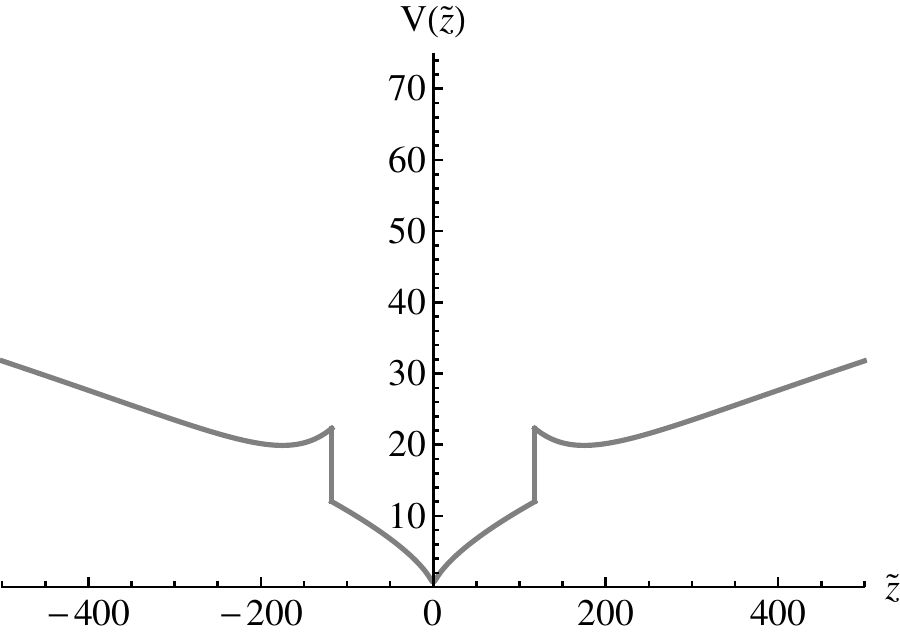}
\caption{Schr\"{o}dinger potential in the antipodal case for density $\tilde{E} = 1000$ in the absence of direct holographic meson-baryon interactions, i.e. $C_{0}(\tilde{z})$ from Eq. \eqref{eq.mesonperturbationsinmedium} encoding the holographic baryons' response to the electric field perturbations is set, by hand, to zero. Comparing this with the second plot within Fig.~\ref{fig.spectrum_antipodal} we see that the emergent potential well gets much more shallow and lacks the barriers present when direct holographic baryon-meson interactions are taken into account. The discontinuity of the potential at the boundary of the charge distribution follows from the discontinuous first derivative of the background radial electric field.}
\label{fig.potential_without_direct_interactions}
\end{figure}

The last point of our analysis is to determine the precise role of the density-enhanced direct holographic meson-baryon interactions on the form of the potential. Figure \ref{fig.potential_without_direct_interactions} shows the Schr\"{o}dinger potential with this effect turned off by hand, and one can clearly see that the most interesting feature present before--i.e., an emergent relatively deep potential well--is now absent. Although our analysis is far from complete, based on our findings we think that capturing the so-far neglected density-enhanced direct holographic meson-baryon interactions is crucial for getting the physics of dense holographic QCD models right. 

To conclude this section, we would also like to stress that in all examples we analyzed we never found states that we could not connect continuously to the ones encountered in the vacuum, as is neatly depicted in Fig. \eqref{fig.spectrum_as_a_func_of_E} for the three lowest vector and axial mesons in the antipodal case. This means that the bound states we found were never genuinely new mesons, but rather those already existing in the vacuum with properties significantly altered by the presence of the dense baryonic medium. If the effect we describe persists in more detailed microscopic studies, then it might have interesting consequences, which we discuss in the last section.

\section{Summary and conclusions \label{sec.summary}}

\subsection{General motivation}
In this paper we have considered the single-flavor Sakai-Sugimoto model at zero temperature and finite baryon density.
Our main motivation was to look for evidence of the existence of the so-called quarkyonic phase \cite{McLerran:2007qj}. 
In the quarkyonic phase, baryons are so tightly packed that quarks no longer belong to a particular baryon; rather, 
one effectively has a quark Fermi surface. Despite this, the excitations of the systems are, as in the vacuum, color singlets, which motivates the name: ``quark'' (from having the quark Fermi surface) and ``yonic'' (from excitations being colorless particles including baryons). The presence of the quark Fermi surface has profound consequences for the properties of the quarkyonic phase: 
chiral symmetry is expected to be restored \cite{McLerran:2007qj}, and
the system is unstable against the formation of quarkyonic chiral spirals \cite{Kojo:2009ha,Kojo:2010fe}. The latter break the chiral symmetry again, together with the translational symmetry, but the chiral condensate oscillates in space around zero with the wavelength related to the chemical potential.

Since discussions of the quarkyonic phase in the QCD literature are phenomenological in nature, it 
is interesting to see to what extent the gauge-gravity duality can shed light on the existence and properties
of this phase. This was the chief motivation for our studies and below we summarize our results detailing once more the assumptions made and listing the most promising open directions.

We began our analysis by motivating (in Sec. \ref{sec.motivationforlargedensity}) why low temperatures and baryon densities of order ${\cal O}(\lambda^{2})$--represented holographically by four-dimensional lattices of the holographic baryons  having extent in both the field theory directions and the holographic direction \cite{Rozali:2007rx}--might correspond to the quarkyonic phase in the holographic model of interest. We listed three reasons that are characteristic of this regime of parameters: the presence of strong electromagnetic interactions between holographically represented baryons, which might be interpreted as a dual manifestation of the formation of a quark Fermi surface \cite{Rozali:2007rx}, the possibility of spontaneous breaking of translational invariance because of instabilities triggered by the Chern-Simons term \cite{Nakamura:2009tf,Ooguri:2010xs}, and density-enhanced interactions of the holographic mesons with baryons, which significantly modify the meson spectrum. Although the first two effects were already anticipated in the holographic literature--albeit never directly in the context of the Quarkyonic phase--the third one, to the best of our knowledge, is a completely new phenomenon.

\subsection{Coarse-grained description of the holographic baryonic medium}

The chief difficulty that studies of very dense holographic QCD are faced with lies in constructing holographic arrays of baryons corresponding to an ${\cal O}(\lambda^{2})$ baryon density in the dual field theory. A fully microscopic solution of the problem would amount to minimizing the energy of a large number of point-like charged particles interacting through DBI electromagnetism, put in a finite volume in the field theory directions and in an external gravitational field associated with the holographic direction. Although construction of such states was already attempted in Ref. \cite{Kaplunovsky:2012gb}, the solutions obtained there consisted of only a few layers of the holographic baryons  in the holographic direction, which is superficially too small a number to observe any of the effects that we mentioned.

Because of these difficulties, following Ref. \cite{Rozali:2007rx}, we adopted a simple coarse-grained description in which we minimized the energy generated by the continuous distribution of the holographic baryons. This way of describing the system turned out to be very similar to the electron star construction of Ref. \cite{Hartnoll:2010gu}, and we drew from this analogy in our considerations, making contact between the two approaches in Sec. \ref{sec.meanfieldattempt}. The effective Lagrangian originally proposed in Ref. \cite{Rozali:2007rx} and used by us is very simple and \emph{postulates} a DBI form of the kinetic term for the mean-field U(1) gauge field and the D8-brane mean-field embedding function and a minimal coupling of the density of the holographic baryons  to the mean-field U(1) gauge field. In the model we also assumed that the energy density of the holographic medium is that of a free system.

All these assumptions might in principle turn out to be violated in a true top-down model of the very dense holographic baryonic medium. In particular, the approach we followed does not take into account oscillations of the microscopic gauge field and the embedding function over the lattice-spacing scale. It is entirely conceivable that such effects will modify the form of the effective Lagrangian, e.g., modify the equation of state or the DBI form of the kinetic term for the mean-field U(1) gauge field. Although we expect the qualitative, rough features of our findings to also hold for truly microscopic charge distributions, quantitative answers might be significantly distorted by the simplifications that were made. This might partly justify promoting the microscopic quantities appearing in the coarse-grained model--like the mass of the holographic baryon and the parameters $\beta$ and $\gamma$ introduced in Sec.~\ref{subsec.solnsforembeddings}--to free parameters in the bottom-up spirit.

\subsection{Antipodal and nonantipodal embeddings}

The original Sakai-Sugimoto model with two D8-branes located antipodally in the circular direction has a generalization, with the asymptotic separation of the D8-branes being an extra parameter and having no direct analogue in QCD. The phenomenology of the Sakai-Sugimoto model does depend on this parameter, denoted here by $L$. For example, the phase diagram is affected by $L$ \cite{Aharony:2006da}, and Yukawa interactions between baryons and scalar mesons \cite{Kaplunovsky:2010eh} depend on $L$ and will vanish in the antipodal case due to the extra $Z_2$ symmetry.

This is the reason why the first new result found in our studies was a generalization of the coarse-grained description of Ref. \cite{Rozali:2007rx} to nonantipodal embeddings. Interestingly, it turned out that within the approximations made by us, the properties of vector and axial mesons in the dense baryonic medium depend rather strongly on $L$, which \emph{a posteriori} is another motivation for this modification of the original Sakai-Sugimoto model.

\subsection{Holographic charge distribution and the quark Fermi surface}

Our coarse-grained description naturally leads to four-dimensional densities of the holographic baryons , as seen also in the more microscopic calculations of Ref. \cite{Kaplunovsky:2012gb}. This is the most distinctive feature of the considered model at ${\cal O}(\lambda^{2})$ baryon densities, as opposed to lower densities, and--together with the common lore of identifying the holographic direction with the energy scale in the dual field theory--one sees that the holographic charge distribution is somewhat reminiscent of a Fermi sea. In fact, it is tempting to directly identify the edge of the holographic baryon distribution with a Fermi surface, as was done by the authors of Ref. \cite{Rozali:2007rx}. Some words of caution are in order here. In the first place--although it is true that in the ground state the lowest lying states that are available to the baryons are being occupied--it is important to remember that it is not the Pauli exclusion principle but rather the Coulomb repulsion which prevents baryons from occupying the same state in the bulk. Moreover,  the states are being occupied in the coordinate (gravity dual) space rather than the momentum space and the baryons are strongly interacting, so a naive weakly coupled second quantized point of view is inappropriate. Having said all that, we think that the best indicator of the presence of a quark Fermi surface is the very strong overlap between the baryons in the dual field theory and (at the same time) the strong interactions between their holographic counterparts. Although the first feature is present at much lower [even ${\cal O}(1)$] baryon densities, the latter arises only for a parametrically dense medium.

\subsection{Chern-Simons-term-induced instabilities}

Apart from the existence of a quark Fermi surface, one of the characteristic features of the quarkyonic phase is the appearance of unstable modes that break translational invariance. In the holographic literature it is known that the presence of the Chern-Simons coupling for electromagnetic/DBI interactions together with the radial electric field may lead to the appearance of unstable modes that break translational invariance \cite{Domokos:2007kt,Nakamura:2009tf,Ooguri:2010kt,Ooguri:2010kt}. So far this mechanism, which appears in a couple of variants \cite{Nakamura:2009tf,Donos:2011bh}, seems to be the only one known within the gauge-gravity duality that might lead to time-independent modulations in holographic media. One of the indications of the instability is the presence of marginally stable modes, i.e., time-independent normalizable perturbations carrying nonzero momentum. A mode of that type, involving a helical modulation in baryon or axial currents, is known to appear in the Sakai-Sugimoto model in the deconfined phase \cite{Ooguri:2010xs}, as well as at zero temperature but large axial chemical potential \cite{Bayona:2011ab}.

As the appearance of this particular mode in the Sakai-Sugimoto model superficially seemed to be rather robust, we performed a thorough search for it for various asymptotic separations of the flavor branes, as well as various charge densities. Details, including interesting technicalities, can be found in Sec. \ref{sec.stability}. It turned out that within the adopted coarse-grained description we could not find such a mode, which we attributed to the smallness of the radial electric field generated by our distributions of the holographic baryons. However, upon promoting the microscopic baryon mass appearing in our coarse-grained Lagrangian to a free parameter, we started seeing marginally stable modes for masses about five times bigger than the microscopic ones, as can be seen in Fig. \ref{fig:MarginalModesDen1000Xi1hPrime3IncreasedMass}. The chief difference between the two models is that the radial electric field generated by the charges on the flavor brane gets a couple of times bigger in the latter case. This might be regarded as an indication that our coarse-grained model needs to be modified in order to capture the physics of interest. It is also important to stress that while searching for instabilities we did not consider the most general ansatz and, hence, there might be unstable modes that evaded our analysis. We leave this issue as an open problem.

Our instability analysis in Sec. \ref{sec.stability} referred to instabilities in the gauge field, whose timelike component is dual to the left and right quark number operators $\psi_L^{\dagger} \psi_L$ and $\psi_R^{\dagger} \psi_R$. Potential instabilities of this type would spontaneously break translational invariance but not affect chiral symmetry. The quarkyonic chiral spiral instability, on the other hand, refers to spatially dependent expectation values for both the chiral condensate as well as operators of the form $\bar{\psi} \gamma_0 \gamma_i \psi$, which also mix left and right quarks. It is possible that a link between the two instabilities exists, and
it would be interesting to explore this further. It might well be that evaluating the chiral condensate in the presence of the condensed Cherm-Simons-term-induced instability gives an expression similar to the one appearing in the quarkyonic chiral spiral.

It is also worth pointing out here that if there is chiral symmetry restoration in the quarkyonic phase, the presence of the quarkyonic chiral spiral would
break the chiral symmetry again but with the condensate averaging to zero over a sufficiently large scale. To reproduce the phenomenology of the quarkyonic phase, we should therefore either find an unstable phase with chiral symmetry restoration, or a stable phase with broken translational
invariance and broken chiral symmetry with the chiral condensate oscillating in space around zero.

\subsection{The spectrum of mesons at parametrically large baryon densities}

As we have emphasized several times, it is only at baryon densities of order ${\cal O} (\lambda^{2})$ that interactions of the flavor-brane gauge-field perturbations with the holographically represented baryons can significantly influence the spectrum of the normalizable modes of the gauge field and hence the masses of the corresponding vector and axial mesons. This happens because the holographic baryons are charged particles, which accelerate in the presence of the electromagnetic field perturbation and in this way lead to a nontrivial dynamical electric current coupled to the DBI electromagnetism on the flavor-brane world volume.

Naively, such an effect should not be visible, as baryons are heavier than their charge by a factor of $\lambda$, and their acceleration due to the electric field associated with the gauge field perturbation and hence their velocities and displacements from equilibrium positions are mass-suppressed. However, the large density of the holographic baryons  compensates the smallness of their velocities and--this is the crucial observation here--the current produced by the holographic baryons yields a contribution to the equation of motion for the gauge field perturbation, which is as important for the dynamics as the gauge field kinetic term. This effect is analogous to the one described by the London equation for electromagnetic waves trying to penetrate a superconductor and can be understood by the effective mass the gauge field acquires inside the medium. This analogy superficially offers a natural mechanism for the chiral (axial) symmetry restoration: a massive gauge field might not be able to penetrate the charge distribution, leading to an effective decoupling of the modes residing on both sides of it.

Although this is what we originally expected to find, the full story turned out to be more complicated. The modifications arise firstly because the gauge field perturbations propagate in a curved spacetime and try to penetrate an inhomogeneous charge distribution, and secondly because the holographic baryons form a lattice and hence one might expect the presence of a restoring force acting on a baryon outside its equilibrium position. The simple analysis in Sec. \eqref{sec.mesons} gave an estimate of the magnitude of the latter effect, and it turned out to contribute, at leading order, at the same order as the kinetic term for the gauge field and as the contribution from the current of the holographic baryons .  Understanding the consequences of this potentially very important effect requires knowledge of the details of the holographic baryon distribution that we clearly lack, and in our analysis we completely neglected it. Our motivation was to show the existence of new phenomena at ${\cal O}(\lambda^{2})$ baryon densities and to motivate a further study, rather than to incorporate all the intricate details right from the outset.

A neat way of understanding the eigenvalue problems of the type encountered here is to write the master equation for perturbations in the form of a Schr\"{o}dinger equation. In this picture the potential appearing in the Schr\"{o}dinger equation allows one to understand the qualitative features of the spectrum obtained through direct calculations.

In the vacuum for the antipodal embedding the potential has the form of a single well that is infinite and supports infinitely many bound states (top plot in Fig. \ref{fig.spectrum_antipodal}). In the nonantipodal case a small substructure emerges, i.e., a small potential barrier whose height, however, does not significantly affect the properties of the mesons (top plot in Fig. \ref{fig.spectrum_nonantipodal}). In particular, for a wide range of asymptotic separations the relative ratio of the masses of the lowest vector and axial vector mesons turns out to be about $40\%$ (Fig. \ref{fig.Deltam}). This can be attributed to the axial symmetry breaking, which is the single-flavor analogue of chiral symmetry breaking in large-$N_{c}$ gauge theories. Put another way, if the chiral symmetry were restored, then the vector and axial mesons would appear in doublets, and in particular would have had equal masses. 

The spectrum of vector and axial mesons turn out to change significantly if we go to ${\cal O}(\lambda^{2})$ densities and include the effects coming from the curved-space analogue of the London equation. The most striking feature that arises is the appearance of an emergent potential well inside the master potential well. The lowest vector and axial mesons then appear as the bound states of the emergent potential rather than the one already existing in the vacuum. This can be very clearly seen in Fig. \ref{fig.spectrum_antipodal}. A phenomenon of this type, to the best of our knowledge, has not yet been encountered in the studies of any of the existing holographic models.

The Schr\"{o}dinger potential at ${\cal O}(\lambda^{2})$ gets even more interesting when we consider nonantipodal embeddings. We already discussed the fact that the emergent potential barrier can support bound states by itself, and that in the nonantipodal embeddings the geometry leads to the appearance of a small potential barrier in the Schr\"{o}dinger potential in the vacuum. At a large enough density these two features combine, leading to an effective potential barrier inside the master potential well. This potential barrier screens the wave functions on both sides and approximately restores the chiral (axial) symmetry between the lowest vector and axial mesons, as can be seen in Fig. \ref{fig.spectrum_nonantipodal}. However, higher states have energies well above the emergent potential barrier and the chiral symmetry cannot be restored for them by means of this mechanism.

It is important to stress that we have found no genuinely new states as a result of our analysis of dense holographic QCD: the axial and vector mesons at ${\cal O}(\lambda^{2})$ baryon densities could be continuously connected to the ones appearing already in the vacuum, as is visible in Fig. \ref{fig.spectrum_as_a_func_of_E}.

Concluding this point, we found evidence for significant modifications of vector and axial meson spectra at baryon densities of order ${\cal O} (\lambda^{2})$. Our analysis was clearly incomplete, as we neglected the restoring force acting on baryons within the holographic lattice.  This might be justified if it turned out that the numerical coefficient associated with this effect was an order of magnitude smaller than the one driving the coupling of the baryon current to the gauge field, but we have no evidence for this. For the moment we do not have any further detailed insight to contribute to this issue. However, the message we want to convey is that there is a set of new phenomena that affect the spectrum of mesons at baryon densities of order ${\cal O} (\lambda^{2})$ that seem to have been overlooked in the literature on this subject, but their precise effect still needs to be worked out in detail. 

\subsection{Chiral condensate and chiral symmetry restoration}

A standard way to examine whether or not the chiral symmetry is broken is to examine whether or not the chiral condensate $\langle \bar{\psi}\psi\rangle=\langle 
{\psi}_L^{\dagger} \psi_R + {\psi}_R^{\dagger} \psi_L \rangle$ vanishes. In the Sakai-Sugimoto model, the left-
and right-chirality quarks sit at different points on the boundary circle, and the gauge-invariant operator--which generalizes the chiral condensate--is $\langle {\psi}_L^{\dagger} P\exp (\int_0^L A_{\tau} d\tau) \psi_R + 
L \leftrightarrow R \rangle$. To compute its expectation value, one needs to compute the regularized area
of a world-sheet disc that ends on the D8-$\overline{\mathrm{D8}}$ branes and on the boundary \cite{Aharony:2008an}.
In the presence of a finite density of baryons, and in the mean-field description that we have been
employing, the computation needs to be somewhat modified. The nontrivial gauge field on
the D8-$\overline{\mathrm{D8}}$ branes couples to the world sheet and will give rise to an extra contribution,
\begin{equation}
{\rm chiral\,\,condensate} \propto P\exp\left(\int_{-\infty}^{+\infty} d\tilde{z} A_{\tilde{z}}\right) \times
{\rm area(disc)} .
\end{equation}
It might appear that the extra contribution vanishes since we chose the gauge $A_{\tilde{z}}=0$. However,
we should evaluate the extra Wilson line in the $\tilde{z}$ direction in the gauge in which all gauge
fields vanish at infinity. One can then see that the extra Wilson line is related to the pure-gauge
behavior of $A_{\mu}$ at infinity in the $A_{\tilde{z}}=0$ gauge. As briefly mentioned in Sec.~\ref{subsec.vacuumsol}, 
and discussed in more detail in Ref. \cite{Sakai:2004cn}, the pure gauge behavior of $A_{\mu}$ encodes the 
expectation value of the pion field, and we can more precisely write
\begin{equation} \label{jjj1}
{\rm chiral\,\,condensate} \propto \exp(2 \Pi(x^{\mu})/f_{\pi}) 
 \times
{\rm area(disc)},
\end{equation}
where $\Pi(x^{\mu})$ is the pion field and $f_{\pi}$ is the pion decay constant. We therefore see
that there is a direct relation between the chiral condensate and the expectation value of the pion field
in our model. In the mean-field description the pion field vanishes and therefore the chiral condensate
does not vanish. In fact, in the antipodal case the geometry of the embedding does not change
as we increase the baryon density, and therefore the chiral condensate does not change either. In the
nonantipodal case the embedding does change, as so does the chiral condensate (but not in a drastic
way), and so chiral symmetry appears to remain broken. We will return to this point below.

We saw that chiral symmetry became less severely broken in the presence
of a finite density but did not quite get restored. It is tempting to speculate about possible qualitative mechanisms
that could cause effective chiral symmetry restoration, as this was one of the main motivations for this work. So
far, the only mechanism for chiral symmetry restoration in these classes of models always involves finite-temperature
physics and a bulk black hole. If the preferred phase involves branes extending all the way to the black hole horizon, 
then chiral symmetry is restored; otherwise, it appears to be broken. So what possible mechanisms could one imagine?

One possibility is that in the presence of the finite-density medium, gauge fields become effectively massive. This effect in our case led to a very intricate form of the Schr\"{o}dinger potential--including in some cases the appearance of a potential barrier associated with the presence of the holographic baryons--but it never caused the gauge field to decay rapidly inside the holographic medium. If this would have been the case, then the ``Fermi surface'' would effectively impose Dirichlet boundary conditions on the gauge field. Consequently the gauge
fields on the two sides of the medium would become decoupled from each other and chiral symmetry would be restored.

A second possibility that exists is that the backreaction becomes important. If the finite-density medium would cause a large 
backreaction on the bulk geometry, causing the geometry to extend further in the IR, then the D8-$\overline{\mathrm{D8}}$ system
could also extend further into the bulk, causing approximate but not exact decoupling between the two ends. However, in
our setup the backreaction does not seem to be important since it is suppressed by $N_f/N_c$ and will become important only at ${\cal O}(N_{c})$ baryon densities.

Another possibility is that chiral symmetry is not exactly restored, but only approximately restored over long distances. Such
a scenario has been proposed in the context of the Skyrme model, where one models the dense nuclear matter using a 
Skyrme crystal (see, e.g., Refs. \cite{Klebanov:1985qi,Goldhaber:1987pb}). In such systems, the Skyrme field $\exp(2 \Pi(x^{\mu})/f_{\pi})$--which also appeared in \cite{Forkel:1989wc}--fluctuates rapidly and averages to zero over longer distances, signaling chiral symmetry restoration. To see similar behavior in our case, we would really need to take the crystal nature of the ground state into account. The closest analogue of the Skyrme lattice in our model is a three-dimensional lattice
of baryons all located at the tip of the D8-brane. As soon as the lattice becomes four-dimensional, the connection
to the Skyrme lattice is less clear. Nevertheless, it would be interesting to investigate whether in our model
a more detailed description of the crystal does give rise to a rapidly fluctuating Skyrme field 
$\exp(2 \Pi(x^{\mu})/f_{\pi})$ on the boundary. If so, one could then also claim--in view of Ref. \cite{Forkel:1989wc}--that the
chiral condensate vanishes when averaged over longer distances. In the Skyrme model, approximate chiral symmetry
restoration is related to the breakup of Skyrmions into half-Skyrmions, and it has been suggested in Ref. \cite{Rho:2009ym}
that this breakup can be translated into a holographic mechanism for chiral symmetry restoration. The precise
implementation of this idea in our setup is, however, not clear to us. 

\subsection{More than one flavor}
The $N_f=1$ case is clearly much simpler than the $N_f>1$ case, where one has to deal with  the non-Abelian world-volume
theory. The $N_f=1$ case is also more singular because there are only point-like instantons in an Abelian theory 
which we modeled using charged particles. For $N_f>1$ one can describe instantons in terms of smooth gauge-field
configurations, though exact multi-instanton solutions in non-Abelian Dirac-Born-Infeld theory are not known. Isolated 
instantons in the $N_f>1$ theory carry an ``orientation'' in flavor space, so for widely spaced instantons one can
roughly think of them in terms of charged particles that carry extra spin-like quantum numbers. Interactions between
instantons crucially depend on their relative orientation \cite{Hashimoto:2009ys,Kaplunovsky:2012gb}, so a mean-field
description would probably be much more complicated and one would need to keep track of the extra quantum numbers. One could
try to get a very crude description of the $N_f>1$ case by taking the $N_f=1$ case and simply adjusting the mass and
charge of the baryons to account for extra contributions to the interactions coming from the orientation. As we saw in Sec. \ref{sec.stability},
such a simple modification of the theory can already cause important qualitative differences. 

\subsection{Relation to other work on the subject}

Earlier works on dense cold Sakai-Sugimoto model include Refs. \cite{Bergman:2007wp,Kim:2007zm,Rozali:2007rx,Kaplunovsky:2012gb}. The crucial difference between our approach and that of Refs. \cite{Bergman:2007wp,Kim:2007zm} is that we take into account the strong electromagnetic repulsion between the holographic baryons arising at ${\cal O} (\lambda^{2})$ densities and study four-dimensional baryon lattices along the lines of Ref. \cite{Rozali:2007rx}. It is fair to say that so far there is no satisfactory macroscopic description of such lattices that starts from first principles. The only microscopic attempt that we are aware of is that of Ref. \cite{Kaplunovsky:2012gb}.

The Chern-Simons-term-induced instabilities in the holographic system are a subject of significant recent research interest, but in the case of holographic QCD there are only a few such papers \cite{Domokos:2007kt,Kim:2010pu,Chuang:2010ku,Ooguri:2010xs,Bayona:2011ab}. The only article among these that considers the baryonic phase at zero temperature in the Sakai-Sugimoto model is Ref. \cite{Chuang:2010ku}, but it focuses on three-dimensional baryon lattices and does not take into account the necessity of checking the normalizability of gauge field perturbations at both ends of the flavor brane, which we discussed in Sec. \ref{sec.stability}.

To the best of our knowledge, the meson spectra at parametrically large densities were studied only in Ref. \cite{Kim:2007zm} (see also Refs. \cite{Kim:2006gp,Kim:2008bv}). This paper also focused on three-dimensional densities of the holographic baryons  and did not take into account the direct holographic meson-baryon interactions that we discussed at length in Sec. \ref{sec.mesons}. The effects taken into account in Ref. \cite{Kim:2007zm} were the nonlinearities of the DBI action, but as shown in Fig. \ref{fig.potential_without_direct_interactions} this is not the key mechanism that affects the masses of mesons. However, one needs to take into account the fact that the radial electric field generated in the three-dimensional approximation to the holographic density of baryons might be bigger than the one in our case for parametrically large densities.

In Ref. \cite{Domokos:2007kt} it was noticed that the Chern-Simons term paired with the radial electric field induces an interesting mixing pattern between the axial and vector mesons carrying nonzero momentum and living in a dense baryonic medium. An interesting spin-off of our work is that the masses of mesons might be significantly different from their vacuum values due to density-enhanced interactions with baryons, which will affect at least the numerical findings of Ref. \cite{Domokos:2007kt} when applied to the Sakai-Sugimoto model.

Recently, an interesting attempt was made to evaluate the chiral condensate at nonzero baryon density \cite{Seki:2012tt}. We want to stress however that this paper focused on three-dimensional densities, and it would interesting to generalize its findings to the four-dimensional densities that we have considered, as well as try to understand possible implications of the granular nature of the holographic medium on the chiral condensate in the dual field theory at nonzero baryon density.

\subsection{Concluding remarks}

We think that the most important outcome of our paper is the realization that parametrically large baryon densities, of order of the square of the 't Hooft coupling constant, may lead to interesting physical effects in the Sakai-Sugimoto model resembling (to some extent) the effects expected to arise in the quarkyonic phase. While some of these effects were already anticipated in the literature--e.g., four-dimensional densities of the holographic baryons appeared already in Ref. \cite{Rozali:2007rx} and were recently discussed in Ref. \cite{Kaplunovsky:2012gb}--some others are new, like the density-enhanced  holographic meson-baryon interactions. The crucial obstacle that we encountered in our studies was the lack of controllable microscopic insight into the nature of the correct coarse-grained description.

Although the main motivation of our work was to make contact with the conjectured quarkyonic phase in large-$N_{c}$ QCD, we found neither the exact chiral  symmetry restoration in the homogeneous phase nor the unstable modes that break translational invariance. It might well be that these features arise once one uses the correct coarse-grained description. An indication of this might be the fact that tweaking the parameters of our effective Lagrangian for perturbations does in fact lead to the appearance of a particular unstable (actually marginally stable) modes.

We think that the most interesting question raised by our studies is the mechanism of the possible chiral symmetry restoration due to a nonzero density. Hence, the most obvious open direction would be to perform a more detailed microscopic analysis of the baryon lattice in order
to understand the fate of the pion at finite densities within these models, as well as to evaluate the chiral condensate more accurately.

\begin{acknowledgements}
We thank Ofer Aharony, Oren Bergman, David Blaschke, Jeffrey Harvey, Vadim Kaplunovsky, Andreas Karch, Keun young Kim, Elias Kiritsis, Gilad Lifschytz, Koenraad Schalm, Cobi Sonnenschein, Shigeki Sugimoto, Alireza Tavanfar and especially David Mateos and Sean Hartnoll for discussions. We also thank Zoltan Fodor, Keun young Kim, Matthew Lippert and Giorgio Torrieri for useful comments on the manuscript. This work was partially supported by the Polish Ministry of Science and Higher Education Grant No. N202 173539, by the Foundation for Fundamental Research on Matter, which is part of the Netherlands Organization for Scientific Research and by the Netherlands Organization for Scientific Research under the NWO Veni scheme (MPH). The work of BDC is supported by the ERC Advanced Grant 268088-EMERGRAV.
\end{acknowledgements}


\onecolumngrid

\appendix
\section{Equations for meson perturbations}
To get the equations for mesonic perturbations we
first consider the generic perturbations 
$\delta \tilde{A}_a \tilde{x}^a=\delta\tilde{a}_3(\tilde{z},\tilde{t})d\tilde{x}^3$
and expand the DBI Lagrangian to second
order in the perturbation,
\begin{equation}
S_{meson} = \int~d\tilde{z}d^4\tilde{x}\Bigg\{\frac{1}{4}\tilde{\xi}^{1/4}(1 + \frac{\tilde{z}^2}{\tilde{\xi}^2})^{1/12}\sqrt{-\det(\tilde{g}_{ab})-\tilde{\xi}^{9/2}(1 + \frac{\tilde{z}^2}{\tilde{\xi}^2})^{3/2}\tilde{A}_0'(\tilde{z})^2}~M^{ab}M^{ch}\tilde{F}_{bc}\tilde{F}_{ha} \Bigg\},
\label{eq:}
\end{equation}
where $M^{ab}=(\tilde{g}_{ab}+\tilde{F}_{ab})^{-1}$
and $\tilde{F}_{ab}$ is the background field strength,
which has the only-nonzero components $\tilde{F}_{z0}=-\tilde{F}_{0z}=\partial_{\tilde{z}}\tilde{A}_0(\tilde{z})$. Using the Euler-Lagrange equations, the direct coupling of 
meson perturbations to the baryon distribution and the ansatz
$\delta\tilde{a}_3(\tilde{z},\tilde{t})= \delta a_3(z)e^{-i\tilde{\omega}\tilde{t}}$,
we get the desired coefficients,

\begin{equation}
C_0^{(\omega)}(\tilde{z}) = 
-\frac{\det  (\tilde{g}_{ab})}{\tilde{\xi }^{5/2} \left(\tilde{\xi }^2+\tilde{z}^2\right)^{5/2}},
\label{eq:}
\end{equation}

\begin{equation}
C_0(\tilde{z}) = \frac{3 \gamma ^2 \tilde{g}^{33} \tilde{\xi }^{3/4} \tilde{w}^2\left(\tilde{z}\right) \left(\tilde{\xi }^2+\tilde{z}^2\right)^{11/12} \sqrt{\tilde{\xi
   }^3+\tilde{\xi } \tilde{z}^2-1} \sqrt{-\det  \tilde{g}_{ab}}}{\beta  \sqrt{9 \sqrt[3]{\tilde{\xi }^2+\tilde{z}^2} \left(\tilde{\xi }^3+\tilde{\xi }
   \tilde{z}^2-1\right) \left(\left(\tilde{\xi }^3+\tilde{\xi } \tilde{z}^2-1\right) \tilde{y}'\left(\tilde{z}\right)^2-\tilde{A}'_0\left(\tilde{z}\right)^2\right)+4
   \tilde{\xi }^{5/3} \tilde{z}^2}}~,
\label{eq:}
\end{equation}

\begin{eqnarray}
C_1(\tilde{z})  & = & \Bigg\{-3 \sqrt[3]{\tilde{z}^2+\tilde{\xi }^2}\left(\tilde{z}^2 \tilde{\xi }+\tilde{\xi }^3-1\right)^2 
\Bigg(-3 (\tilde{z}^2+\tilde{\xi }^2)
\tilde{A}_0'(\tilde{z}) \tilde{A}_0''(\tilde{z})+\\
\nonumber
& + &5\tilde{z} \tilde{A}_0'(\tilde{z})^2+\tilde{y}'(\tilde{z})
(\tilde{z} (5-2 \tilde{\xi } (\tilde{z}^2+\tilde{\xi }^2)) \tilde{y}'(\tilde{z})+3 (\tilde{z}^2+\tilde{\xi }^2)
   (\tilde{z}^2 \tilde{\xi }+\tilde{\xi }^3-1) \tilde{y}''(\tilde{z}))\Bigg)\\
\nonumber
&-&4 \tilde{z}\tilde{\xi }^{5/3} (-2 \tilde{z}^4
  \tilde{\xi }-\tilde{z}^2 (\tilde{\xi }^3-1)+\tilde{\xi }^5-\tilde{\xi }^2)\Bigg\}
\\
\nonumber
& / &\left(\tilde{z}^2+\tilde{\xi }^2\right) \left(\tilde{z}^2 \tilde{\xi }+\tilde{\xi }^3-1\right) \left(9 \sqrt[3]{\tilde{z}^2+\tilde{\xi }^2} \left(\tilde{z}^2
   \tilde{\xi }+\tilde{\xi }^3-1\right) \left(\left(\tilde{z}^2 \tilde{\xi }+\tilde{\xi }^3-1\right)
   \tilde{y}'\left(\tilde{z}\right)^2-\tilde{A}_0'\left(\tilde{z}\right)^2\right)+4 \tilde{z}^2 \tilde{\xi }^{5/3}\right).
\label{eq:}
\end{eqnarray}

The density of baryons on the D8-branes takes a simple algebraic form and is given by\begin{eqnarray}
\tilde{w}^2(\tilde{z})  &=&
{\tilde{\xi }^3 \left(\tilde{\xi
   }^2+\tilde{z}^2\right)^{4/3}}\left[D_1(\tilde{z})\tilde{A}'_0(\tilde{z})+D_2(\tilde{z})\tilde{A}''_0(\tilde{z})+D_3(\tilde{z})\tilde{A}'^3_0(\tilde{z})\right]
\\
\nonumber
&/&\Bigg\{81 \gamma  \sqrt{-\det(\tilde{g}_{ab})} \left(\tilde{\xi }^3+\tilde{\xi } \tilde{z}^2-1\right)^2 \left(-\tilde{\xi }^{3/2} \left(\tilde{\xi
   }^2+\tilde{z}^2\right)^{3/2} \tilde{A}_0'\left(\tilde{z}\right)^2-\det(\tilde{g}_{ab})\right)^{3/2}\Bigg\},
\label{eq:}
\end{eqnarray}
where the coefficients are
\begin{eqnarray}
D_1(\tilde{z})  &=& 9 \sqrt[3]{\tilde{\xi }} \Bigg(3 \left(\tilde{\xi }^2+\tilde{z}^2\right) \left(\tilde{\xi }^3+\tilde{\xi } \tilde{z}^2-1\right)^2 \tilde{y}'\left(\tilde{z}\right)\Bigg(\tilde{z} \left(2 \tilde{\xi } \left(\tilde{\xi }^2+\tilde{z}^2\right)-5\right) \tilde{y}'\left(\tilde{z}\right)\\
\nonumber
&-&3 \left(\tilde{\xi }^2+\tilde{z}^2\right)
   \left(\tilde{\xi }^3+\tilde{\xi } \tilde{z}^2-1\right) \tilde{y}''\left(\tilde{z}\right)\Bigg)
   + 4 \tilde{\xi }^{5/3} \tilde{z} \left(\tilde{\xi
   }^2+\tilde{z}^2\right)^{2/3} \left(-\tilde{\xi }^5+\tilde{\xi }^2+2 \tilde{\xi } \tilde{z}^4+\left(\tilde{\xi }^3-1\right) \tilde{z}^2\right)\Bigg),
\label{eq:}
\end{eqnarray}
\begin{eqnarray}
D_2(\tilde{z}) = 9 \sqrt[3]{\tilde{\xi }} \left(\tilde{\xi }^2+\tilde{z}^2\right) \left(\tilde{\xi }^3+\tilde{\xi } \tilde{z}^2-1\right) \left(9 \left(\tilde{\xi
   }^2+\tilde{z}^2\right) \left(\tilde{\xi }^3+\tilde{\xi } \tilde{z}^2-1\right)^2 \tilde{y}'\left(\tilde{z}\right)^2+4 \tilde{\xi }^{5/3} \tilde{z}^2
   \left(\tilde{\xi }^2+\tilde{z}^2\right)^{2/3}\right),
\label{eq:}
\end{eqnarray}

\begin{equation}
D_3(\tilde{z}) = -135 \sqrt[3]{\tilde{\xi }} \tilde{z} \left(\tilde{\xi }^2+\tilde{z}^2\right) \left(\tilde{\xi }^3+\tilde{\xi } \tilde{z}^2-1\right)^2.
\label{eq:}
\end{equation}

The determinant of the metric is

\begin{equation}
\sqrt{-\det (\tilde{g}_{ab})} = \frac{\tilde{\xi }^{3/4} \left(\tilde{\xi }^2+\tilde{z}^2\right)^{7/12} \sqrt{9 \sqrt[3]{\tilde{\xi }^2+\tilde{z}^2} \left(\tilde{\xi }^3+\tilde{\xi }
   \tilde{z}^2-1\right)^2 \tilde{y}'\left(\tilde{z}\right)^2+4 \tilde{\xi }^{5/3} \tilde{z}^2}}{3 \sqrt{\tilde{\xi }^3+\tilde{\xi } \tilde{z}^2-1}}.
\label{eq:}
\end{equation}

The embedding function, obtained
from the equations of motion, is given by

\begin{equation}
\tilde{y}'(\tilde{z}) = 
\frac{\tilde{\xi }^{5/3} \sqrt{\tilde{\xi }^3-1} \sqrt{4 \tilde{z}^2 \tilde{\xi }^{5/3}-9 \sqrt[3]{\tilde{z}^2+\tilde{\xi }^2} \left(\tilde{z}^2 \tilde{\xi
   }+\tilde{\xi }^3-1\right) \tilde{A}_0'\left(\tilde{z}\right)^2}}{3 \sqrt[6]{\tilde{z}^2+\tilde{\xi }^2} \left(\tilde{z}^2 \tilde{\xi }+\tilde{\xi }^3-1\right)
   \sqrt{\left(\tilde{z}^2+\tilde{\xi }^2\right)^{5/3} \left(\tilde{z}^2 \tilde{\xi }+\tilde{\xi }^3-1\right)-\tilde{\xi }^{19/3}+\tilde{\xi }^{10/3}}}.
\label{eq:}
\end{equation}

As was explained in the text it is
explicitly seen that the embedding 
function depends on the baryon density
only through the background gauge field (and the matching condition on the boundary of the charge distribution).
In the antipodal case $\tilde{\xi}=1$ 
and $\tilde{y}'(\tilde{z})=0$, which makes the shape of the D8-brane independent of the 
baryon density.

\twocolumngrid

\bibliography{cold_biblio}{}
\bibliographystyle{utphys}

\end{document}